\documentclass[prc,aps,floats,floatfix,showpacs,nofootinbib,%
superscriptaddress]{revtex4}

\usepackage{amsmath}
\usepackage{amssymb}
\usepackage{amsfonts}
\usepackage{graphicx}
\usepackage{tabularx}
\usepackage{multirow}
\usepackage{lscape}
\usepackage{rotating}
\usepackage{pstricks}
\usepackage{feynmf}
\usepackage{lscape}
\usepackage{feynmf}

\begin{document}

\title{Impact of statistical uncertainties on the composition of the outer crust of a neutron star}

\author{A. Pastore}
\email{alessandro.pastore@york.ac.uk}
\affiliation{Department of Physics, University of York, Heslington, York, Y010 5DD, UK}

\author{D. Neill}
\affiliation{Department of Physics, University of York, Heslington, York, Y010 5DD, UK}

\author{H. Powell}
\affiliation{Department of Physics, University of York, Heslington, York, Y010 5DD, UK}

\author{K. Medler}
\affiliation{Department of Physics, University of York, Heslington, York, Y010 5DD, UK}

\author{C. Barton}
\email{charles.barton@york.ac.uk}
\affiliation{Department of Physics, University of York, Heslington, York, Y010 5DD, UK}


\begin{abstract}
\end{abstract}


 
\date{\today}
\begin{abstract}
By means of Monte Carlo methods, we perform a full error analysis on the Duflo-Zucker mass model. In particular, we study the presence of correlations in the residuals to obtain a more realistic estimate of the error bars on the predicted binding energies. To further reduce the discrepancies between model prediction and experimental data we also apply a Multilayer Perceptron Neural Network. We show that the root mean square of the model further reduces of roughly 40\%.
We then use the resulting models to predict the composition of the outer crust of a non accreting neutron star. We provide a first estimate of the impact of error propagation on the resulting equation of state of the system.
\end{abstract}


\maketitle

\section{Introduction}
The study of neutron stars (NS)  is important for our understanding of nuclear matter in the most extreme conditions.  The recent measurements of gravitational waves~\cite{abbott2017gw170817} followed by optical detection of electromagnetic radiation ~\cite{abbott2017multi} has opened an entire new way of studying such objects. 

Due to the large pressure gradient, matter within NS arranges in layers with different  properties. Going from the outside toward the inside (low density to high density), we find the \emph{envelope}: a thin layer of iron atoms. At densities of $\rho_b\approx 10^4$ $g/cm^3$, the outer crust begins where a gas of (ultra) relativistic  electrons surrounds ionised nuclei. 
Throughout the \emph{outer crust} the Fermi energy of the electrons increases with density thus making it energetically favourable for electron capture to occur. As a consequence, we observe an increasing neutronisation of matter leading to the production of very neutron rich nuclei. The process continues until the drip-line is reached and neutrons start dripping off to form a uniform gas~\cite{pastore2011superfluid,pastore2012superfluid,chamel2012neutron,pearson2012inner,pastoreDRIP,gezerlis2014pairing,pastore2017new}. This region is called the \emph{inner crust}.
By going deeper inside the NS, the density continues increasing and eventually nuclei dissolve into a Fermi liquid, the \emph{core}. 
The composition of the core is still matter of open debate and several models have been suggested over years~\cite{alford2007astrophysics,chamel2008physics,sharma2015unified,chatterjee2016hyperons,vidana2018d}.

 
 In this article, we focus our attention on the sequence of nuclei forming the outer crust and how a full fledged error analysis impacts the prediction of the equation of state (EoS).
To determine the sequence of nuclei which constitute the outer crust, we follow the methodology developed by Baym, Pethick and Sutherland (BPS) in 1971~\cite{baym1971ground}. For each input pressure value, one obtains the number of neutrons ($N$) and protons ($Z$) that minimises the Gibbs free energy for a given lattice configuration.

To perform these calculations a key ingredient is the nuclear binding energy. Of course, when experimental measurements are available there is no (or little uncertainty) in doing these calculation. The problem arises when one needs to use binding energies extrapolated from mass models. How reliable are these values?
Only recently  the scientific community has tried to answer such a question by performing detailed error analysis and error propagation on models~\cite{dobaczewski2014error,schunck2015error,mcdonnell2015uncertainty,neufcourt2018bayesian} and use them for astrophysical purposes~\cite{utama2017refining,utama2018validating,sprouse2019propagation}. 

At present a very limited set of models are published with relevant statistical information to perform a complete error analysis~\cite{kortelainen2015propagation,roca2015covariance,becker2018error}, and none of these has been fully developed for astrophysical purposes. Of course, we may argue that a model which is very accurate in reproducing known masses may be very bad when used to extrapolate values and vice-versa. In this work, we decided to study one of the widely used mass-models currently used in astrophysical calculations, namely the Duflo-Zucker (DZ) mass model~\cite{duflo1995microscopic} and to perform a full statistical analysis.

The DZ model describes nuclear masses of known nuclei with a small standard deviation of $\approx 500$ keV, in its 10 parameter version (DZ10) and $\approx 400$ keV in its 33 parameter version (DZ33). The DZ represents an ideal case for our statistical investigations for two main reasons: the parameters have been adjusted only on one type of observable and it is computationally inexpensive.
By means of Monte Carlo (MC)  methods we have thus performed a full error propagation and studied the impact of these error bars on the resulting sequence of nuclei in the outer crust of a NS.
To further improve the quality of the predictions, we equip the DZ with a Neural Network (NN) adjusted on  mass residuals. We use such a combined model to predict the composition of the outer crust of NS. 

 The article is organised as follows: in Sec.\ref{sec:DZ}, the values and errors in the parameters of the Duflo-Zucker model are calculated and propagated into the binding energies calculated by the model. In Sec.\ref{BNN}, we briefly present the main features of the NN and its application to the DZ model. The BPS method for calculating the composition of the outer crust is detailed in Sec.\ref{sec:Outer}. In this section we present the impact of error bars on nuclear masses on the possible composition of the outer crust. Finally we present our conclusions in Sec.\ref{sec:conclusion}.


\section{Duflo-Zucker mass model}\label{sec:DZ}

The Duflo-Zucker \cite{duflo1995microscopic}  is a very successful macroscopic mass model based on a generalised liquid-drop (LD) plus  the shell-model monopole Hamiltonian and It is used to obtain the binding energies of nuclei along the whole nuclear chart with high accuracy.
Following Refs~\cite{mendoza2010anatomy,zuker2011anatomy,qi2015theoretical}, we write the nuclear binding energy ($B$) for a given nucleus as a sum of ten terms as

\begin{eqnarray}\label{bene}
B_{th}=a_1V_C+a_2 (M+S)-a_3\frac{M}{\rho}-a_4V_T+a_5V_{TS}+a_6s_3-a_7\frac{s_3}{\rho}+a_8s_4+a_{9}d_4+a_{10}V_P\;.
\end{eqnarray}

\noindent We defined $A=N+Z$ ; $2T=|N-Z|$  and $\rho=A^{1/3}\left[ 1-\frac{1}{4}\left(\frac{T}{A}\right)^2\right]^2$.
The ten different contributions can be grouped in two categories: in the first one we find terms similar to LD model as Coulomb ($V_C$), symmetry energy ( $V_T,V_{TS}$) and pairing $V_P$. The other parameters originate from the averaging of shell-model Hamiltonian and they are based on the microscopic single level structure of the nucleus. For a more detailed discussion we refer to Ref.~\cite{mendoza2010anatomy}, where all these terms are described in great detail. Within the scientific literature, it is possible to find other variants of the DZ model containing 28 and 33 terms~\cite{qi2015theoretical}. The DZ28 and DZ33 have a different structure compared to Eq.\ref{bene} and we refer to Ref~\cite{duflo1995microscopic} for more details. In the present article, we will consider only  the DZ10 model as given in Eq.\ref{bene}.

Following the methodology illustrated in Ref.~\cite{pastore2019introduction}, we fit DZ10 using the nuclear mass table of 2016~\cite{wang2017ame2016}. We define a penalty function of the form

\begin{eqnarray}\label{chi2}
\chi^2=\sum_{N,Z\in\text{data-set}}\frac{\left[B_{exp}(N,Z)-B_{th}(N,Z)\right]^2}{\sigma^2(N,Z)}\;.
\end{eqnarray}

In the fit, we  have excluded all nuclei with $A<14$; moreover we have considered only measured binding energies  with an experimental error lower than $100$ keV. Ignoring the small experimental uncertainties, we set $\sigma^2(N,Z)=1$, thus assuming equal weights to all data.
In total we considered 2293 nuclei. The optimal parameter set $\mathbf{a}^0=\{a_1,a_2,a_3,\dots,a_{10} \}$ is found using standard Nelder Mead method~\cite{vetterling1994numerical} and it is given in Tab.\ref{tab:data}. The DZ10 values are compatible with previous analysis illustrated in Ref.~\cite{mendoza2010anatomy} and based on 1995 mass table~\cite{audi19951995} and more recent Ref.~\cite{uta17} based on 2012 mass table~\cite{wang2012ame2012}. In the second  and third column of the table, we report the errors on the parameters named type-1 and type-2 obtained by calculating the Hessian matrix of the $\chi^2$ at its minimum and using the Block-Bootstrap method illustrated in Sec. \ref{sec:boot}.

\begin{table}[!h]
\begin{center}
\begin{tabular}{c|ccc}
\hline
\hline
 &     Value                   & Error  & Error \\
 &                        &   type-1  &  type-2\\
\hline
$a_1$ & 0.70454     & 0.00037 & 0.0010 \\
$a_2$ &   17.740     &0.0068  &0.020 \\
$a_3$ &   16.244& 0.023& 0.069 \\
$a_4$ &   37.500   & 0.042 &0.119 \\
$a_5$ &   53.56     & 0.20 & 0.53 \\
$a_6$ &  0.4573     & 0.0069 &0.018 \\
$a_7$ &   2.071     & 0.035 & 0.093\\
$a_8$ &   0.0210 & 0.0002  & 0.0004\\
 $a_9$ &  41.43      & 0.21  &0.66 \\
 $a_{10}$ &  6.162 &0.088   & 0.103\\
\hline
\hline
\end{tabular}
\label{tab:data}
\caption{Parameters of the DZ10 mass formula given in Eq.\ref{bene}. The quantities are expressed in MeV. The last two columns represent the error on the parameters neglecting/considering correlations among residuals. See text for details.}
\end{center}
\end{table}

The basic assumption used in fitting a model is that it satisfies the following equation
\begin{eqnarray}\label{eq:hyp}
B_{exp}(N,Z)=B_{th}(N,Z|\mathbf{a}^0)+\varepsilon(N,Z)\;,
\end{eqnarray}

\noindent where $\varepsilon(N,Z)$ represents the residuals. In an ideal case, $\varepsilon(N,Z)$ follow a normal distribution with zero mean and variance $\sigma^2$.

\begin{figure}[!h]
\begin{center}
\includegraphics[width=0.38\textwidth,angle=-90]{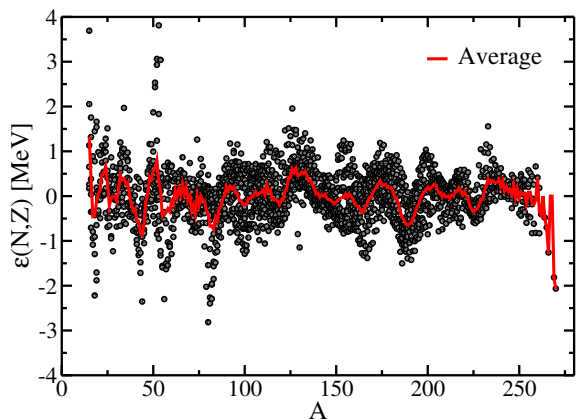}
\end{center}
\caption{Residuals of the DZ10 model. The solid line represents an average as given in Eq.\ref{varia}.}
\label{residual}
\end{figure}

In Fig.\ref{residual}, we report $\varepsilon(N,Z)$ as a function of the mass number A. The root mean square deviation is $\sigma=0.572$ keV. Using the data shown in Fig.\ref{residual}, we have performed a simple Kolmogorov-Smirnov (KS) test to check whether or not the residuals follow a normal distribution. We obtained a very low $p$-value thus rejecting the null-hypothesis of normally distributed data~\cite{bar89}.
Following Ref.~\cite{bertsch2017estimating}, we define an average residual for each value of $A$ as

\begin{eqnarray}\label{varia}
\sigma^2_A=\frac{1}{N_A}\sum_{Z+N=A}\left( \mathcal{E}(N,Z)-\mathcal{E}_A(A)\right)^2\;.
\end{eqnarray}

\noindent where $\mathcal{E}_A(A)$ is the average value of the residuals for the nuclei with mass $A$. The averaging is reported in Fig.\ref{residual} as a solid line. Although we loose some information in the $(N-Z)$ direction, $\sigma^2_A$ still carries some useful information concerning the trend along the A direction of the nuclear chart.

\begin{figure}[!h]
\begin{center}
\includegraphics[width=0.38\textwidth,angle=0]{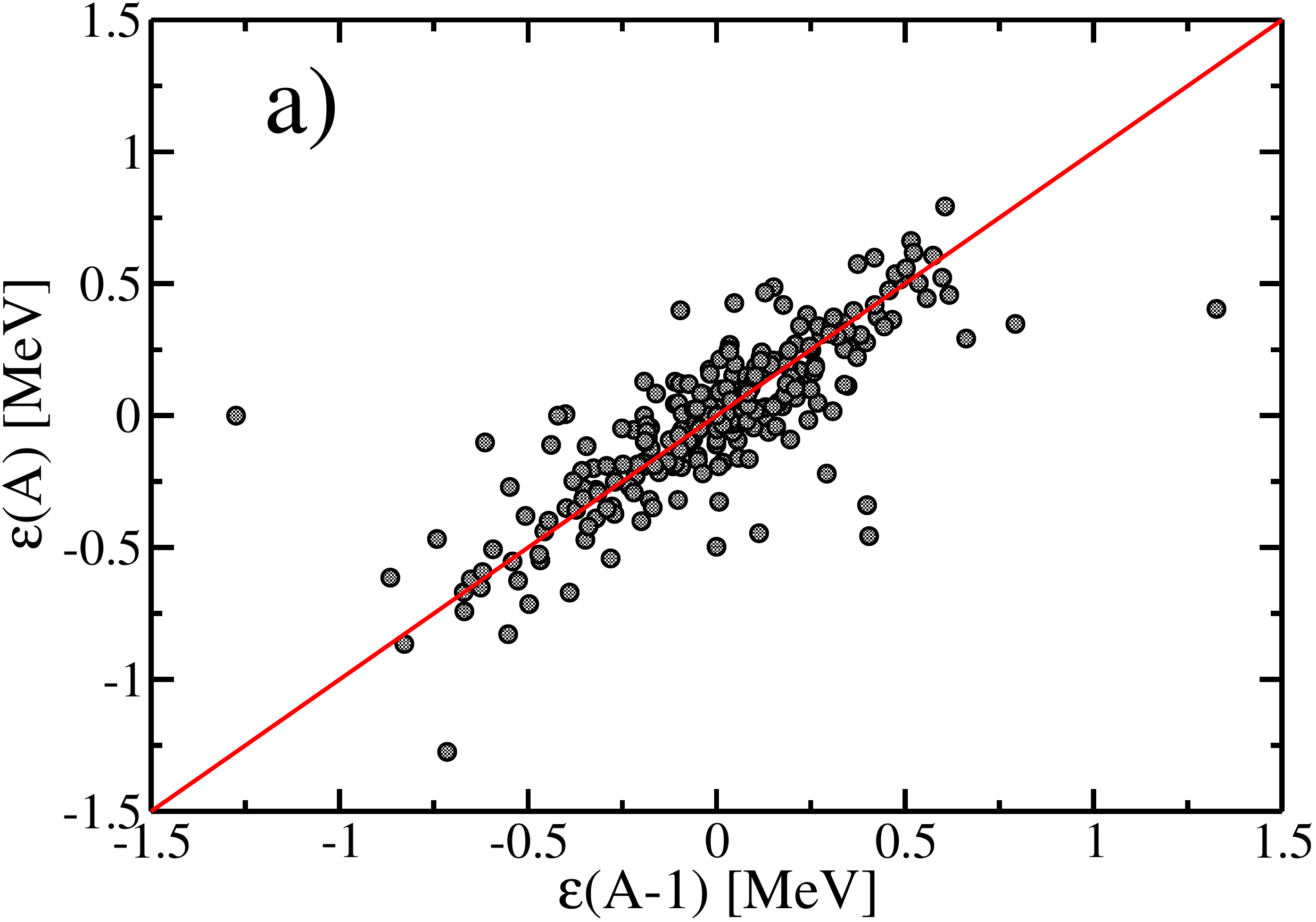}
\includegraphics[width=0.38\textwidth,angle=0]{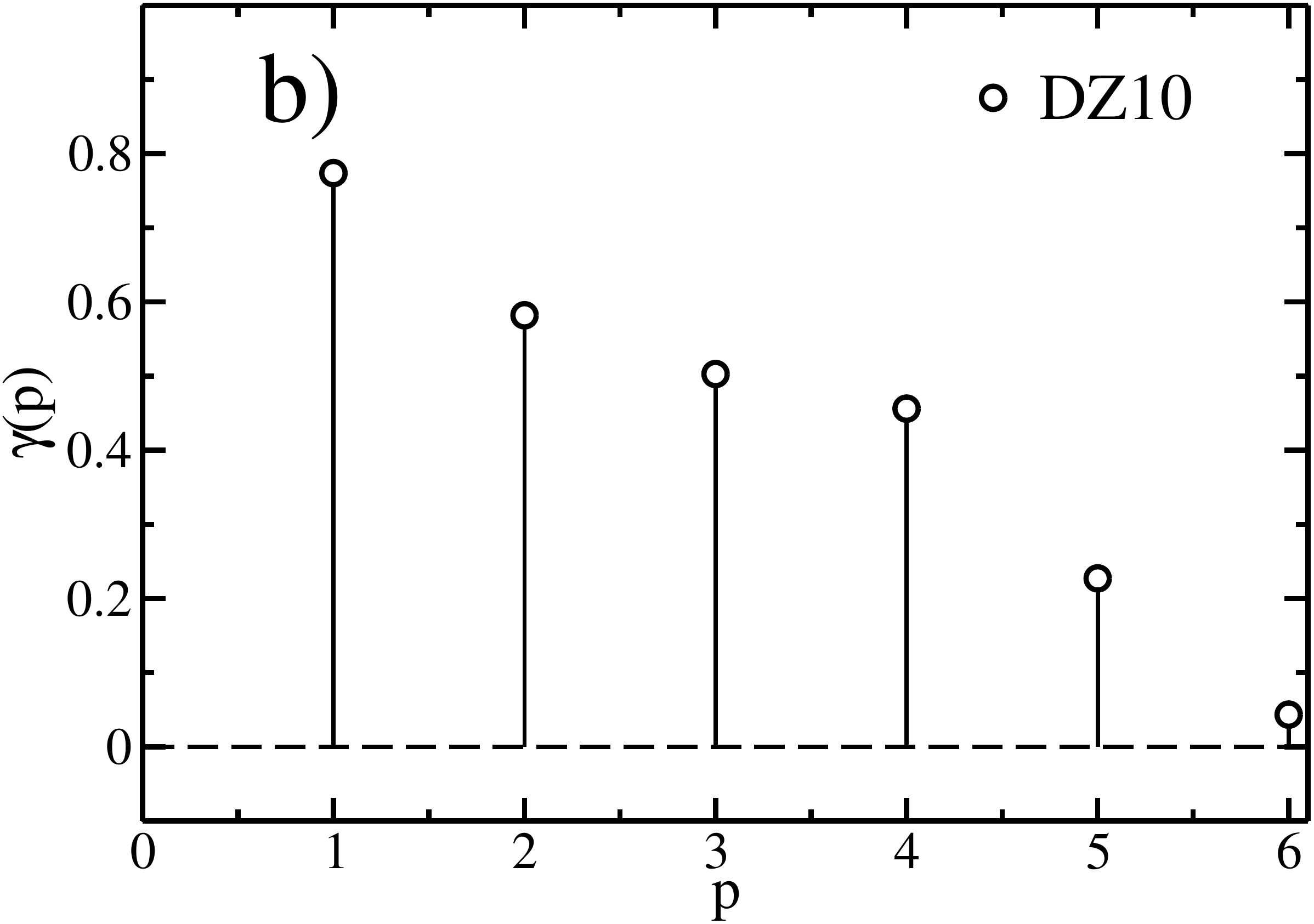}
\end{center}
\caption{(Colour online). Left panel: lag-plot of order 1 for DZ10 model. Right panel: evolution of the self-correlation coefficient defined in Eq.\ref{selfcor} for different values of the lag. See text for details.}
\label{lag}
\end{figure}

We  use the lag-plot  of lag-$p$ ~\cite{brockwell2013time} to determine the degree of correlation between data. The  lag-plot  consists in plotting the data against themselves, but shifted by $p$ units.
In case of non-correlated residuals, the lag-plot does not have any pattern, \emph{i.e} one should observe a cloud of points. In Fig. \ref{lag} (panel a), we show the lag-1 plot of the averaged residuals. They cluster around the diagonal, meaning a strong auto-correlation. To be more quantitative, we also define a self-correlation coefficient as

\begin{eqnarray}\label{selfcor}
\gamma(p)=\frac{\sum_i (X_i-\bar{X}) (Y_i-\bar{Y})}{\sigma_X\sigma_Y}\;,
\end{eqnarray}

\noindent where $Y_i=X_{i-p}$ is the delayed series.
In Fig.\ref{lag}b, we present the values of $\gamma(p)$  for different lags. We observe that there is a non-negligible correlation  (i.e. $\gamma(p) \ge 0.5$) for DZ10  up to $p\approx4$; this means that the residual of a nucleus $A$ is strongly correlated with all other residuals within the interval $[A-4,A+4]$.

These correlations should be taken into account to provide a more realistic estimate of the error of the parameters. See detailed discussion in Refs~\cite{bertsch2017estimating,pastore2019introduction}. A simple approach to perform a more consistent error analysis is based on Block Bootstrap (BB)~\cite{pastore2019introduction}  and it discussed in the following sections.

\subsection{Error propagation}\label{sec:err:porp}

Leaving aside for a moment the problem of  correlations in the residuals as discussed in Fig.\ref{lag}, we now focus on the propagation of errors in the DZ model on the binding energies.
For the present analysis, we will use the standard method based on a truncated Taylor expansion of $\chi^2$ function around its minimum

\begin{eqnarray}\label{error1}
\sigma_{th}^2=\mathcal{G}\mathcal{C} \mathcal{G}^T\;,
\end{eqnarray}

\noindent where $\mathcal{G}$ is the vector containing the derivative of the model in parameter space

\begin{eqnarray}
\mathcal{G}_{j}=\frac{\partial B_{th}}{\partial a_j}\;.
\end{eqnarray}

The latter are done numerically using the prescription given in Ref.~\cite{roca2015covariance}. The covariance matrix $\mathcal{C}$ is obtained by evaluating the Hessian matrix of Eq.\ref{chi2} at its minimum. We name this method of evaluating errors type-1 to distinguish from the Block Bootstrap one, see Tab.\ref{tab:data}.
Having  access to errors, we compare how the DZ10  performs against experimental data.We  expect that 68\% of known masses  differ from the model prediction no more than $\sigma=\sigma_{th}+\sigma_{exp}$, where $\sigma_{exp}$ is the experimental error reported in Ref.~\cite{wang2017ame2016}. By increasing the error bar by a factor of 2 and 3 we should obtain 95\% and 99.7\% of experimental binding energies falling into the interval.
In Tab.\ref{tab:percentage}, we have reported the actual percentage of nuclei using type-1 errors  falling within the error band for the full nuclear chart and some different mass regions.

\begin{table}[!h]
\begin{center}
\begin{tabular}{c|ccc}
\hline
\hline
                       \multicolumn{4}{c}{type-1 errors}  \\
 \hline
                        & $1\sigma$ & $2\sigma$ & $3\sigma$ \\
\hline
Full chart & 13.6\% &  27.2\% & 39.5\%    \\
$50\le  A <150$ & 14.7\% & 26.8\% & 37.2\%  \\
$20 \le Z \le 50 $  &11.5 \% &  22.2\% &  31.4\%  \\
$A \ge 150 $  & 14.8\% & 30.8 \% &  45.8\%  \\
\hline
\hline
\end{tabular}
\label{tab:percentage}
\caption{Percentage of nuclei included in the total error bars in different sectors of the nuclear chart. See text for details. }
\end{center}
\end{table}

Focusing on the nuclei relevant for the NS crust composition, \emph{i.e.} $20 \le Z \le 50 $, we observe that only 11.5\% of measured nuclei fall within the error bars of DZ model. This means that either Eq.\ref{eq:hyp} is not valid and one should improve the model or the estimate of the error bars is not done properly.
In the next section we discuss how to improve error bars taking into account residual correlations, while in  Sec.\ref{BNN}, we discuss how to improve the model using a neural network.

\subsection{Block Bootstrap}\label{sec:boot}

Bootstrap is a simple Monte Carlo  procedure commonly used to estimate error bars of various statistical estimators~\cite{efron1994introduction,kreiss2011bootstrap,chernick2011bootstrap,muir2018bootstrap}. Firstly introduced by Efron in the 70s ~\cite{efr79}, it has become a very common statistical tool in several scientific domains~\cite{ras87,fisher1990new,sauerbrei1992bootstrap,manly2006randomization,per14,pasq18}

The crucial aspect of non-parametric bootstrap (NPB) is to generate a new series of residuals $\mathcal{E}^*(A)$ from random sampling the original one $\mathcal{E}(A)$ and use them to generate a new set of observables (in this case the binding energy) as

\begin{eqnarray}
B^*(N,Z)=B_{exp}(N,Z)+\mathcal{E}^*(N,Z)\;.
\end{eqnarray}

\noindent  $B^*(N,Z)$ is used as input for a least square minimisation. As discussed in Ref.~\cite{pastore2019introduction}, NPB gives access to the full likelihood of the model from which one can easily extract confidence intervals on the estimated parameters.  

Another important advantage of Bootstrap is the possibility of removing the usual approximation of Eq.\ref{error1} and to use a full MC error propagation instead. The latter is a side product of any NPB procedure since effectively, we produce several DZ fits and thus we can infer the error on each predicted mass by simply defining the 68\% quantile on each mass distribution.
In the current case, we did not find any noticeable difference between the full MC error propagation and simpler approximation provided in Eq. \ref{error1}.

A crucial step to improve error bar estimates is to take into account correlations in the residuals. A possible way would be to model a non diagonal variance matrix entering the minimisation procedure of the $\chi^2$ function given in Eq.\ref{chi2}.  In the present article, we stick to Bootstrap methods, but using the Block-Bootstrap (BB) variant ~\cite{kreiss2012bootstrap}.
The data are separated in overlapping blocks to which we apply the Monte Carlo procedure discussed before. In such a way, we preserve the correlations among residuals of neighbouring nuclei.

Our data set presents correlations along the $A$ and $(N-Z)$ direction. In Refs~\cite{bertsch2017estimating,pastore2019introduction}, the length of correlations in the $A$ directions was more important that the $(N-Z)$ one; as a consequence it has been possible to simplify the problem by taking the average value and correct it by a Gaussian error extracted from the isotopic dependent variance defined in Eq.\ref{varia}. From  Fig.\ref{lag}, we observe that the length of correlations along the $A$ direction is quite short, few units, thus the approximation done before in Ref. ~\cite{bertsch2017estimating,pastore2019introduction} is not \emph{a priori} justified. 
We have thus decided to develop a more consistent framework for error analysis using a 2-dimensional BB.  The main issue here is that the residuals represent an irregular grid. We have  created square blocks of size $l$ around each nucleus using the following procedure: when less than half terms ($l^2$) are empty, since there are no data, we discard such a block. Otherwise we keep it.
Finally we inspect in all retained blocks for the presence of missing entries, these are replaced by a random number extracted from a normal distribution with the variance equal to the one of the DZ10 model.

\begin{figure}[!h]
\begin{center}
\includegraphics[width=0.38\textwidth,angle=-90]{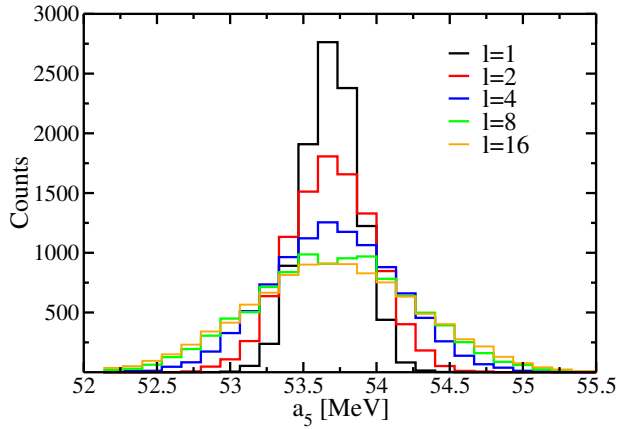}
\end{center}
\caption{(Colour online). Histogram of the $a_5$ parameter, Eq.\ref{bene}, as a function of the block size. See text for details.}
\label{boot:histo}
\end{figure}

The choice of $l$ is somehow arbitrary, but in the present case, by  using the lag-plot shown in Fig.\ref{lag}, we can assess that a block size of $l=8$ will take into account all correlations. 
 In Fig.\ref{boot:histo}, we illustrate the histogram of the $a_5$ parameter as a function of the block size for $N_{boot}=5000$ Monte Carlo samplings. We see that beyond $l=8$ the increase in the size of the blocks does not lead to any major change in the distribution of the values of $a_5$.
As discussed in Refs~\cite{bertsch2017estimating,pastore2019introduction}, correlations do not change the mean value of the distribution, but they modify the width. As a consequence the resulting error bars become larger.

\noindent From the marginalised likelihood, we extract the full correlation matrix $\mathcal{R}$, defined as

\begin{eqnarray}
\mathcal{R}_{ij}=\frac{\mathcal{C}_{ij}}{\sqrt{\mathcal{C}_{ii}}\mathcal{C}_{jj}}\;.
\end{eqnarray}

\begin{figure}[!h]
\begin{center}
\includegraphics[width=0.38\textwidth,angle=0]{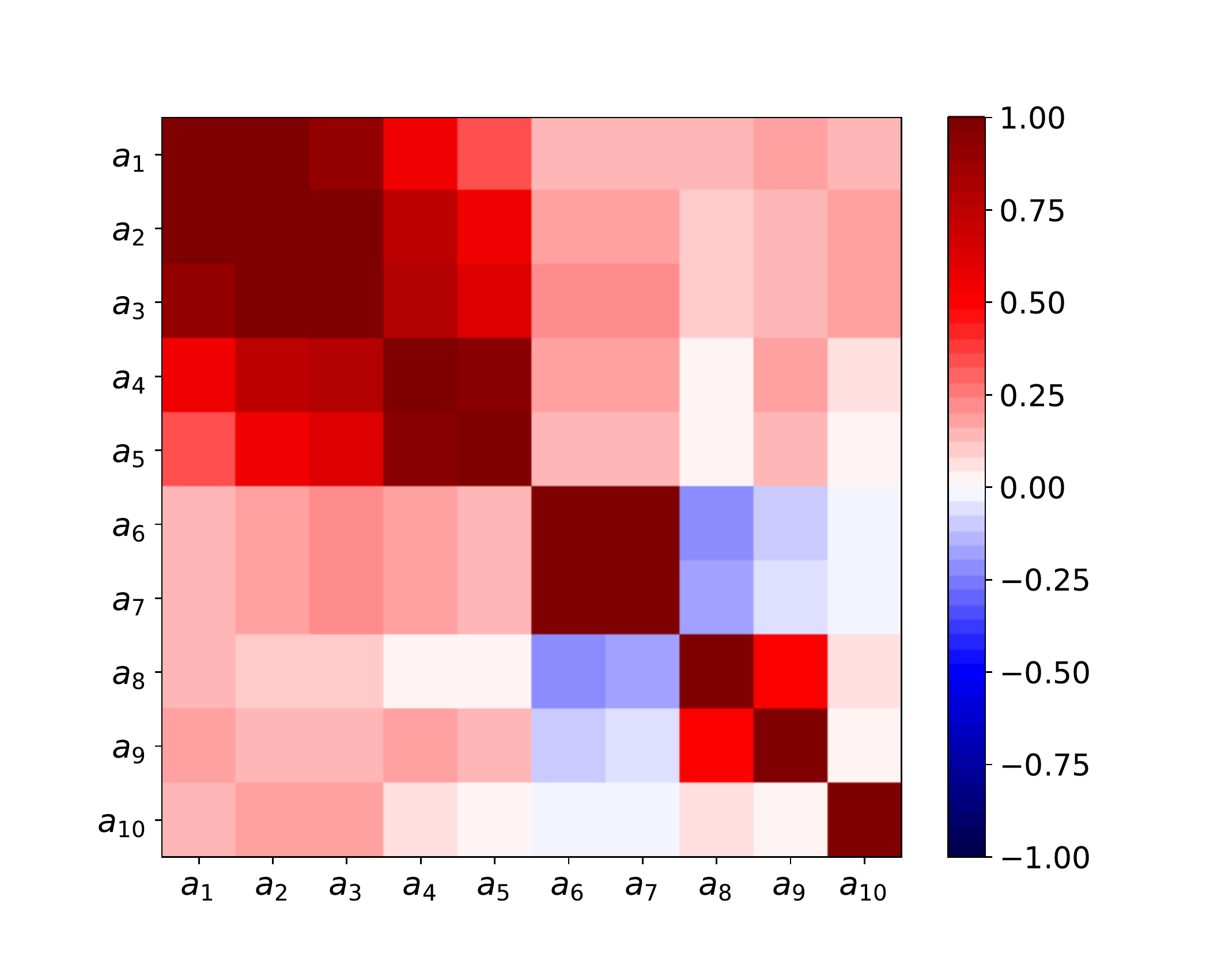}
\end{center}
\caption{(Colour online).Correlation matrix obtained using BB. See text for details.}
\label{cov}
\end{figure}

\noindent  By construction, the matrix elements of $\mathcal{R}$ are normalised between -1 and 1. The resulting matrix for the DZ10 model is reported in Fig.\ref{cov}.
We observe a strong correlation between the two \emph{master} terms $a_2,a_3$ and Coulomb coupling $a_1$. We also notice a degree of correlation between these three terms and the two related to symmetry energy $a_4,a_5$. The coupling constants $a_6,a_7$ show a very strong correlation between them. Most likely it could be possible to express $a_7$ as a function of $a_6$ and remove one of these parameters~\cite{nikvsic2016sloppy}.
Finally, the $a_8,a_9$ terms are mildly correlated to each other and completely uncorrelated to the other surface terms.
The last term $a_{10}$ takes into account the pairing effects. It is completely uncorrelated to all terms, thus showing that pairing correlations within DZ10 are absorbed only by $a_{10}$.

\begin{table}[!h]
\begin{center}
\begin{tabular}{c|ccc}
\hline
\hline
    \multicolumn{4}{c}{type-2 errors}  \\
 \hline
                        & $1\sigma$ & $2\sigma$ & $3\sigma$ \\
\hline 
Full chart & 34.5\% &  60.4\% & 77.9\%    \\  
$50\le  A \le 150$ & 31.8\% & 55.5\% & 74.2\%  \\
$20 \le Z \le 50 $  &27.9 \% &  52.8\% &  71.9\%  \\
$A > 150 $  & 39.9\% & 69.4 \% &  85.6\%  \\
\hline
\hline
\end{tabular}
\label{tab:percentage2}
\caption{Percentage of nuclei included in the total error bars using BB error bars for different sectors of the nuclear chart. }
\end{center}
\end{table}

We  repeat the comparison between theoretical and experimental values using the new error bars. The results are reported in Tab.\ref{tab:percentage2}.
 We see that the current estimate perform much better than the one based on type-1 errors. The percentage of nuclei falling into the error bars is not still the expected value (\emph{i.e.} 68\% of counts for $1\,\sigma$ error), although the error bars are probably more realistic than in the previous case.
Given the discrepancy of the results observed in Tab.\ref{tab:percentage2}, we can conclude, also observing Fig.\ref{residual}, that to improve the reproduction of the data, we need to act on the model itself. To this purpose we employ a Neural Network.

\section{Multilayer Perceptron}\label{BNN}

To capture the fluctuating function needed to accurately predict the binding energy of nuclei, we used a class of feedforward artificial neural network (NN) called  Multilayer Perceptron (MLP)~\cite{gurney2014introduction,svozil1997introduction}. A NN is a system of connected algorithms, called nodes, that are designed to mimic the working of a biological brain. These nodes are arranged into layers and we allow for connections only between nodes of two sequential layers, with the input nodes having no node connected behind them and the output nodes not connecting to any further nodes.

Following the work of Ref~\cite{uta17}, we built a NN composed by two layers densely connected having 45 and 84 nodes each. We use a sigmoid function as activation and stochastic gradient method to determine the values of the weights of each neural connection. 

We train the NN over the residual shown in Fig.\ref{residual}. To avoid over-fitting, we split the set in a training set (2/3 of the data) and a control set. The splitting is done randomly. By evolving several epochs we have trained the network until the loss function, defined as the distance between the output of the NN and the data, reaches a plateau~\cite{murata1994network}.

We then replace in Eq.\ref{eq:hyp}, the new theoretical model composed by the DZ10 plus the NN. 
We can compare the performance of such a model by comparing with the nuclear masses. We obtain a root mean square deviation of $\sigma_{DZ+NN}=0.344$ keV. This means that the NN has been able to grasp a signal in the residual and thus further improve the agreement with experimental data.
Such a result is in agreement with previous findings ~\cite{niu2018nuclear,utama2017refining}.
Several architecture of the NN are possible, but from our analysis we did not obtain any significant improvement.

\begin{figure}[!h]
\begin{center}
\includegraphics[width=0.38\textwidth,angle=0]{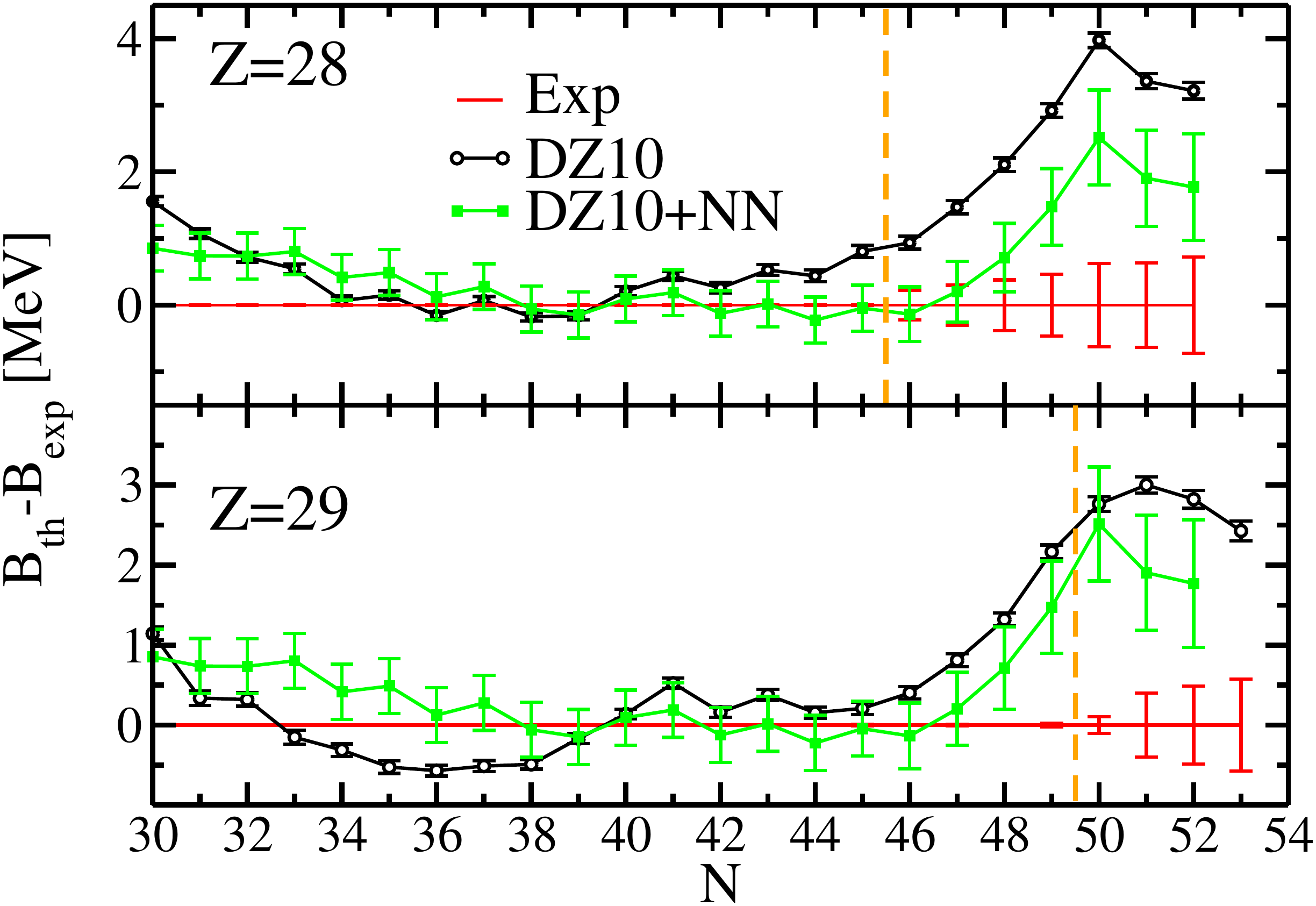}
\end{center}
\caption{(Colour online) In the two panel we compare the evolution of the difference between the theoretical and experimental binding energies obtained using only a DZ10 model (open dots) and a DZ10 plus NN (full squares). 
On the horizontal line we also report the experimental error on measured masses. The vertical dashed line separates masses that have been directly measured from estimated ones. }
\label{expmass}
\end{figure}

In Fig.\ref{expmass}, we illustrate the evolution of the difference between the theoretical binding energies and the experimental one~\cite{wang2017ame2016,welker2017binding} obtained using the DZ10 (open dots) and DZ10+NN(full squares). On the figure, we also provide the theoretical and experimental error bar. The typical error bars on masses are not visible on the energy scale, apart from the estimated values of masses. These values are separated from the measured one by a vertical dashed line. 
The theoretical bar associated with the DZ10 has been obtained using BB method, while for the DZ10+NN model, we have defined the error bar by adding in quadrature the error of the original DZ10 model (type-2) and the variance of the NN over the nuclear chart. This is not the optimal way to estimate the error and a more sophisticated analysis is necessary.
In general, we notice that the inclusion of NN reduces the discrepancy between the theory and the nuclear masses, thus showing that the DZ10 is missing some physical effect. Further studies are still necessary to draw solid conclusions, but we have notice that even using more powerful NN it is hard to further reduce such a discrepancy~\cite{idini2019pac}. This may be explain in terms of chaotic motion inside the nucleus as discussed in Ref.\cite{barea2005nuclear}. Further studies are thus necessary.

\section {Outer crust}\label{sec:Outer}

To determine the chemical composition of the outer crust, we follow the methodology illustrated in Ref.~\cite{roca2008impact}, \emph{i.e.} we determine the optimal combination of neutrons and protons that minimises the baryonic chemical potential $\mu_b$ at a given pressure $P$

\begin{equation}
\mu_b=\frac{W_N+\frac{4}{3}W_L+Z\mu_e}{A}\;.
\end{equation}

\noindent  The three terms appearing in this formula are the lattice energy $W_L$, the nuclear energy $W_N$ and the electron chemical potential $\mu_e$.
The minimisation is done assuming charge neutrality and $\beta$-stable matter. At zero temperature the Gibbs free energy is related to the total energy of the system as $\mu_b=G/A$. 

Following Ref.\cite{baym1971ground}, the outer crust is formed by a body-centre-cubic (BCC) lattice of nuclei surrounded by free electrons~\cite{coldwell1960zero}. The energy contribution of the lattice reads

\begin{equation}
W_L=-1.819620\frac{Z^2e^2}{4\pi \varepsilon_0a},
\end{equation}

\noindent where $a$ is the lattice constant which is given by $n_Na^3=2$. The density of the nuclei is related to the electronic density by the relation

\begin{equation}
n_N=\frac{n_e}{Z}.
\end{equation}

\noindent The energy contribution coming from finite nuclei reads

\begin{equation}
W_N=m_n(A-Z)+m_pZ-B(N,Z)\;.
\end{equation}

\noindent Finite nuclei do not contribute to the total pressure, which is just the sum of the lattice $P_L$ and electron contribution $P_e$
\begin{eqnarray}P&=&P_e+P_L\\
&=&n_e\mu_e-E_e+\frac{1}{3}W_L\frac{n_e}{Z}\;,
\end{eqnarray}

\noindent where $\mu_e=\frac{\partial E_e}{\partial n_e}$ and $E_e$ is the total energy of the electrons. In our case, this is just the kinetic term

\begin{eqnarray}
E_e=\frac{m_e^4}{8\pi^2 n_e}\left[ x_F y_F (x_F^2+y_F^2)-\ln(x_F+y_F)\right]\;,
\end{eqnarray}

\noindent where $x_F=p_{Fe}/m_e$ and $y_F=\sqrt{1+x_F^2}$.

The full minimisation using the DZ mass-model leads to the result shown in Tab.\ref{tab:DZ}.  When the value of the binding energy is known experimentally, we read such a value from  AME2016 mass table~\cite{wang2017ame2016}, while if it is estimated or absent we use the DZ10 model to produce the binding energy.
As already observed by other authors~\cite{baym1971ground,ruster2006outer,pearson2011properties,sharma2015unified} we see that the outer crust of a non-accreting neutron star is composed (almost independently of the model used) by semi-magic nuclei.
From Tab.\ref{tab:DZ}, we observe that at low pressure region, nuclei with N=50 are favoured and at  $P\approx$ 10$^{-4}$ MeVfm$^{-3}$  the N=82 nuclei appear.

\begin{table}[!h]
\begin{center}
\begin{tabular}{cccc}
\hline
\hline
 $P_{max}$ [MeVfm$^{-3}$] &$\rho^{max}_b$ [fm$^{-3}$] &N &Z  \\
\hline
    3.30 $\cdot 10^{-10}$   &  4.67 $\cdot 10^{-6}$     &30 &26  \\
   4.36$\cdot 10^{-8}$  &   1.52 $\cdot 10^{-4}$    &34 &28 \\
 3.56$\cdot10^{-7}$  &   7.46 $\cdot 10^{-4}$      & 36   &       28\\
    4.02$\cdot10^{-7}$     &  8.57 $\cdot 10^{-4}$      &    38       &   28  \\
   1.03$\cdot10^{-6}$   &   1.73 $\cdot 10^{-3}$   &      50   &       36  \\
   5.59$\cdot10^{-6}$  &   6.56 $\cdot 10^{-3}$      & 50      &    34   \\
                  \hline
   1.76$\cdot10^{-5}$&   1.54 $\cdot 10^{-2}$  &          50     &     32    \\
         1.58$\cdot10^{-4}$  &   8.94 $\cdot 10^{-2}$    &   50 &         28 \\
   1.82$\cdot10^{-4}$   &1.03 $\cdot 10^{-1}$ &    82        &  42   \\
   3.31$\cdot10^{-4}$   &1.73 $\cdot 10^{-1}$  &      82       &   40  \\
   4.83$\cdot10^{-4}$  &2.29 $\cdot 10^{-1}$ &     82         & 38    \\
   4.86$\cdot10^{-4}$&2.39 $\cdot 10^{-1}$&          82         & 36   \\
      \hline
      \hline
\end{tabular}
\label{tab:DZ}
\caption{Composition of the outer crust of a NS using the DZ mass model.  In the first  two columns we report the maximum value of pressure and baryonic density at which the nucleus is found using the minimisation procedure. The horizontal line separates the measured mass reported in AME2016~\cite{wang2017ame2016}. }
\end{center}
\end{table}

From Tab.\ref{tab:DZ}, we see that thanks to major experimental effort several nuclei in the N=50 region relevant for the NS have been measured with extreme accuracy ~\cite{welker2017binding}, but still the masses of nuclei in the region N=82 need to be predicted using  some particular mass model~\cite{sobiczewski2014predictive}.
Following the discussion in Sec.\ref{BNN}, we have also performed a full minimisation of the outer crust using the DZ10+NN. The result is reported in Tab.\ref{tab:DZNN}

\begin{table}[!h]
\begin{center}
\begin{tabular}{cccc}
\hline
\hline
 $P_{max}$ [MeVfm$^{-3}$] &$\rho^{max}_b$ [fm$^{-3}$] &N &Z  \\
\hline
    3.30 $\cdot 10^{-10}$   &  4.67 $\cdot 10^{-6}$     &30 &26  \\
   4.36$\cdot 10^{-8}$  &   1.52 $\cdot 10^{-4}$    &34 &28 \\
 3.56$\cdot10^{-7}$  &   7.46 $\cdot 10^{-4}$      & 36   &       28\\
    4.02$\cdot10^{-7}$     &  8.57 $\cdot 10^{-4}$      &    38       &   28  \\
   1.03$\cdot10^{-6}$   &   1.73 $\cdot 10^{-3}$   &      50   &       36  \\
   5.59$\cdot10^{-6}$  &   6.56 $\cdot 10^{-3}$      & 50      &    34   \\
                  \hline
   1.76$\cdot10^{-5}$&   1.54 $\cdot 10^{-2}$  &          50     &     32    \\
      2.67$\cdot10^{-5}$&   2.19 $\cdot 10^{-2}$  &          50     &     30    \\
         1.45$\cdot10^{-4}$  &   8.15 $\cdot 10^{-2}$    &   50 &         28 \\
   1.84$\cdot10^{-4}$   &1.04 $\cdot 10^{-1}$ &    82        &  42   \\
   3.33$\cdot10^{-4}$   &1.68 $\cdot 10^{-1}$  &      82       &   40  \\
   4.83$\cdot10^{-4}$  &2.29 $\cdot 10^{-1}$ &     82         & 38    \\
      \hline
      \hline
\end{tabular}
\label{tab:DZNN}
\caption{Composition of the outer crust of a NS using the DZ10+NN mass model.  In the first  two columns we report the maximum value of pressure and baryonic density at which the nucleus is found using the minimisation procedure. The horizontal line separates the measured mass reported in AME2016~\cite{wang2017ame2016}. }
\end{center}
\end{table}

The inclusion of NN in the mass models has an impact over the general composition. It affects both the threshold pressures and the observed sequence of nuclei  in the crust. The major difference between the results reported in Tab.\ref{tab:DZ} and \ref{tab:DZNN} is the absence (presence)  of $^{80}$Zr and the presence (absence) of $^{118}$Kr at the  transition between outer and inner crust. 

\subsection{Error propagation in the outer crust}\label{sec:MC}

\begin{figure}[!h]
\begin{center}
\includegraphics[width=0.38\textwidth,angle=0]{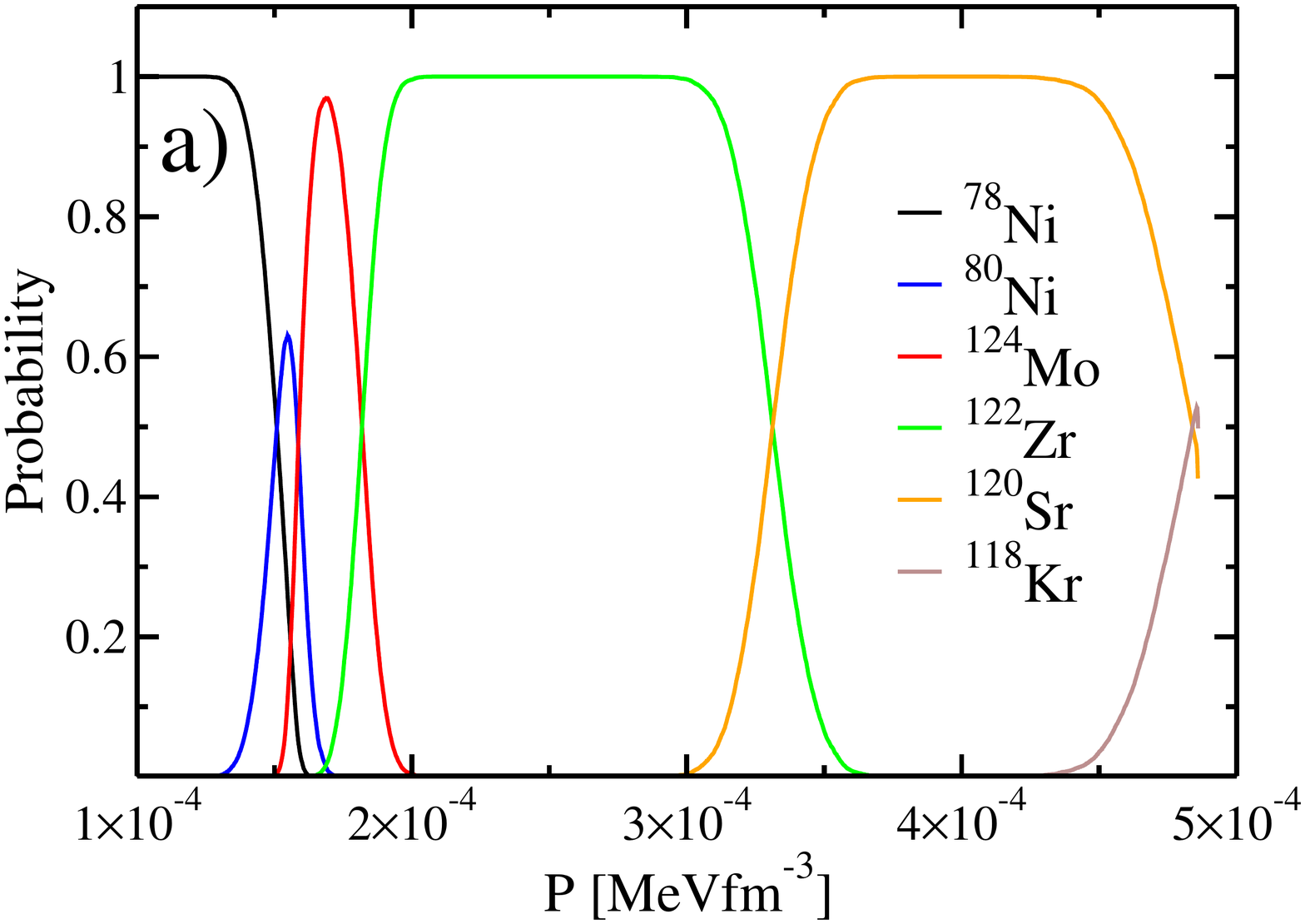}
\includegraphics[width=0.38\textwidth,angle=0]{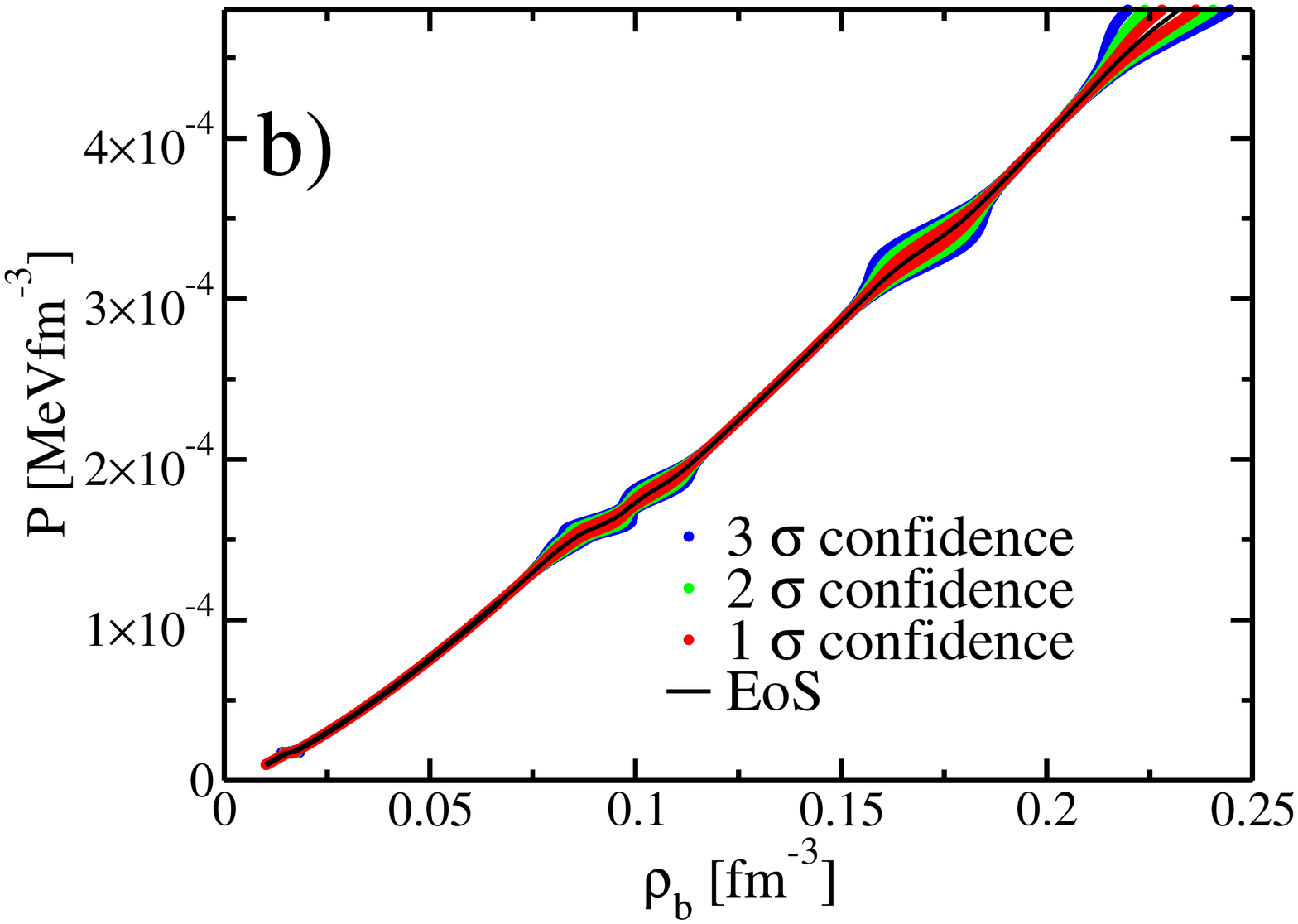}
\end{center}
\caption{(Colour online) Panel a: existence probability of a given nucleus  within the outer crust  as a function of the pressure. On panel b, the equation of state of the system including statistical uncertainties. See text for details.}
\label{probability}
\end{figure}

Having studied in detail the error propagation during the fitting procedure, we now assess the impact of such errors on the composition of the outer crust. The case of type-1 errors have been already presented in Ref.~\cite{neill2019impact}, here we apply the same methodology  to the case of type-2 errors only. 
To this purpose we build 10$^4$ new mass tables~\cite{kirson2012empirical} defined as

\begin{eqnarray}
B^*_{th}(N,Z)=B_{th}(N,Z)+\mathcal{N}(0,\sigma(N,Z))\;,
\end{eqnarray} 

\noindent where $B_{th}(N,Z)$ is the value predicted by DZ10 model and $\mathcal{N}(0,\sigma(N,Z))$ represents a random normal error using as a variance the error on masses discussed in Sec.\ref{sec:boot}.
When a mass is known experimentally, we  use the experimental value. When the mass is not know or estimated, we use the DZ10 model. Having access to such a large number of samples, we define an \emph{existence} probability for each nucleus as a function of pressure within the DZ model. The probability is defined in the usual \emph{frequentist} approach~\cite{bar89} as the number of successful event over the total number of events.
In Fig.\ref{probability} (a), we present the probability of appearance of a given nucleus as a function of the pressure of the star. The explicit inclusion of error bars on nuclear masses does not change the main sequence reported in Tab.\ref{tab:DZ}, but we observe the transitional region between N=50 and N=82 are no more sharp.
Another interesting aspect is that $^{80}$Zr is not included in the main sequence obtained with DZ10, but when considering error bars, we have a probability of 60\% of observing it. 

Using the same data set, we can define in a similar way a statistical uncertainty for the EoS: by counting the 10$^4$ EoS built before, we define the 68\%, 95\% and 99\% quantile of the counts, \emph{i.e}, one, two or three $\sigma$ deviation. The result is reported in Fig.\ref{probability} (b). We interpret the bands as statical uncertainties propagated from our model. As expected, we observe quite large uncertainties corresponding to the densities at which the chemical composition changes. By using these error bars, it is now possible to compare the EoS obtained with various models and check if the results are compatible or not within the statistical errors.

Finally, it would be interesting to observe how such uncertainties impact other observables on NS as for example radii~\cite{fortin2016neutron}. We leave such an aspect for future investigations.

\section{Conclusions}\label{sec:conclusion}

Astrophysical models strongly rely on nuclear data and even more on estimates based on theoretical models. In the current article, we have performed a full statistical analysis of a widely used mass model DZ10. We have shown that the standard approximation of adopting independent residuals is not correct, moreover to assess the quality of extrapolation based on the global variance of the model it is only a crude approximation. 
We have  illustrated how using Block Bootstrap, it is possible to identify correlations among the parameters of the model and to provide a more \emph{realistic} estimate of its error bars. We have then used such error bars to propagate the statistical uncertainties on the equation of state of the outer crust of a neutron star.
We have observed that the most favourable nuclear configuration are the semi-magic one with N=50 and N=82. The presence of error impacts strongly the transition between these two magic number. We have provided a prediction in terms of probabilities of the most favourable nuclei to happen. By applying such a systematic method to various mass model, we aim at providing a robust guidance for future experimental measurements to identify key nuclei to improve our knowledge of the outer crust of a neutron star.

Finally, we have also studied the use of neural networks to further reduce the discrepancy between the DZ10 model and the experimental masses.  Although a  more systematic approach is required~\cite{neufcourt2018bayesian}, we have obtained very encouraging results, in good agreement with previous findings~\cite{utama2017refining}. The use of NN allows for a further reduction of the global root mean square deviation of the model to $\sigma_{DZ+NN}=0.344$ keV, a value which is quite close to the most accurate mass model available in the literature~\cite{min11}.


\section*{Acknowledgements}

This work has been supported by STFC Grant No. ST/P003885/1.
The authors want to thanks M. Mougeot for interesting discussions about current experimental measurements of nuclear binding energies.

\bibliography{biblio}

\begin{thebibliography}{67}
\expandafter\ifx\csname natexlab\endcsname\relax\def\natexlab#1{#1}\fi
\expandafter\ifx\csname bibnamefont\endcsname\relax
  \def\bibnamefont#1{#1}\fi
\expandafter\ifx\csname bibfnamefont\endcsname\relax
  \def\bibfnamefont#1{#1}\fi
\expandafter\ifx\csname citenamefont\endcsname\relax
  \def\citenamefont#1{#1}\fi
\expandafter\ifx\csname url\endcsname\relax
  \def\url#1{\texttt{#1}}\fi
\expandafter\ifx\csname urlprefix\endcsname\relax\def\urlprefix{URL }\fi
\providecommand{\bibinfo}[2]{#2}
\providecommand{\eprint}[2][]{\url{#2}}

\bibitem[{\citenamefont{Abbott et~al.}(2017{\natexlab{a}})\citenamefont{Abbott,
  Abbott, Abbott, Acernese, Ackley, Adams, Adams, Addesso, Adhikari, Adya
  et~al.}}]{abbott2017gw170817}
\bibinfo{author}{\bibfnamefont{B.~P.} \bibnamefont{Abbott}},
  \bibinfo{author}{\bibfnamefont{R.}~\bibnamefont{Abbott}},
  \bibinfo{author}{\bibfnamefont{T.}~\bibnamefont{Abbott}},
  \bibinfo{author}{\bibfnamefont{F.}~\bibnamefont{Acernese}},
  \bibinfo{author}{\bibfnamefont{K.}~\bibnamefont{Ackley}},
  \bibinfo{author}{\bibfnamefont{C.}~\bibnamefont{Adams}},
  \bibinfo{author}{\bibfnamefont{T.}~\bibnamefont{Adams}},
  \bibinfo{author}{\bibfnamefont{P.}~\bibnamefont{Addesso}},
  \bibinfo{author}{\bibfnamefont{R.}~\bibnamefont{Adhikari}},
  \bibinfo{author}{\bibfnamefont{V.}~\bibnamefont{Adya}}, \bibnamefont{et~al.},
  \bibinfo{journal}{Physical Review Letters} \textbf{\bibinfo{volume}{119}},
  \bibinfo{pages}{161101} (\bibinfo{year}{2017}{\natexlab{a}}).

\bibitem[{\citenamefont{Abbott et~al.}(2017{\natexlab{b}})\citenamefont{Abbott,
  Abbott, Abbott, Acernese, Ackley, Adams, Adams, Addesso, Adhikari, Adya
  et~al.}}]{abbott2017multi}
\bibinfo{author}{\bibfnamefont{B.~P.} \bibnamefont{Abbott}},
  \bibinfo{author}{\bibfnamefont{R.}~\bibnamefont{Abbott}},
  \bibinfo{author}{\bibfnamefont{T.}~\bibnamefont{Abbott}},
  \bibinfo{author}{\bibfnamefont{F.}~\bibnamefont{Acernese}},
  \bibinfo{author}{\bibfnamefont{K.}~\bibnamefont{Ackley}},
  \bibinfo{author}{\bibfnamefont{C.}~\bibnamefont{Adams}},
  \bibinfo{author}{\bibfnamefont{T.}~\bibnamefont{Adams}},
  \bibinfo{author}{\bibfnamefont{P.}~\bibnamefont{Addesso}},
  \bibinfo{author}{\bibfnamefont{R.}~\bibnamefont{Adhikari}},
  \bibinfo{author}{\bibfnamefont{V.}~\bibnamefont{Adya}}, \bibnamefont{et~al.},
  \bibinfo{journal}{Astrophys. J. Lett} \textbf{\bibinfo{volume}{848}},
  \bibinfo{pages}{L12} (\bibinfo{year}{2017}{\natexlab{b}}).

\bibitem[{\citenamefont{Pastore et~al.}(2011)\citenamefont{Pastore, Baroni, and
  Losa}}]{pastore2011superfluid}
\bibinfo{author}{\bibfnamefont{A.}~\bibnamefont{Pastore}},
  \bibinfo{author}{\bibfnamefont{S.}~\bibnamefont{Baroni}}, \bibnamefont{and}
  \bibinfo{author}{\bibfnamefont{C.}~\bibnamefont{Losa}},
  \bibinfo{journal}{Physical Review C} \textbf{\bibinfo{volume}{84}},
  \bibinfo{pages}{065807} (\bibinfo{year}{2011}).

\bibitem[{\citenamefont{Pastore}(2012)}]{pastore2012superfluid}
\bibinfo{author}{\bibfnamefont{A.}~\bibnamefont{Pastore}},
  \bibinfo{journal}{Physical Review C} \textbf{\bibinfo{volume}{86}},
  \bibinfo{pages}{065802} (\bibinfo{year}{2012}).

\bibitem[{\citenamefont{Chamel}(2012)}]{chamel2012neutron}
\bibinfo{author}{\bibfnamefont{N.}~\bibnamefont{Chamel}},
  \bibinfo{journal}{Physical Review C} \textbf{\bibinfo{volume}{85}},
  \bibinfo{pages}{035801} (\bibinfo{year}{2012}).

\bibitem[{\citenamefont{Pearson et~al.}(2012)\citenamefont{Pearson, Chamel,
  Goriely, and Ducoin}}]{pearson2012inner}
\bibinfo{author}{\bibfnamefont{J.}~\bibnamefont{Pearson}},
  \bibinfo{author}{\bibfnamefont{N.}~\bibnamefont{Chamel}},
  \bibinfo{author}{\bibfnamefont{S.}~\bibnamefont{Goriely}}, \bibnamefont{and}
  \bibinfo{author}{\bibfnamefont{C.}~\bibnamefont{Ducoin}},
  \bibinfo{journal}{Physical Review C} \textbf{\bibinfo{volume}{85}},
  \bibinfo{pages}{065803} (\bibinfo{year}{2012}).

\bibitem[{\citenamefont{Pastore et~al.}(2013)\citenamefont{Pastore, Margueron,
  Schuck, and Vi\~nas}}]{pastoreDRIP}
\bibinfo{author}{\bibfnamefont{A.}~\bibnamefont{Pastore}},
  \bibinfo{author}{\bibfnamefont{J.}~\bibnamefont{Margueron}},
  \bibinfo{author}{\bibfnamefont{P.}~\bibnamefont{Schuck}}, \bibnamefont{and}
  \bibinfo{author}{\bibfnamefont{X.}~\bibnamefont{Vi\~nas}},
  \bibinfo{journal}{Phys. Rev. C} \textbf{\bibinfo{volume}{88}},
  \bibinfo{pages}{034314} (\bibinfo{year}{2013}),
  \urlprefix\url{https://link.aps.org/doi/10.1103/PhysRevC.88.034314}.

\bibitem[{\citenamefont{Gezerlis et~al.}(2014)\citenamefont{Gezerlis, Pethick,
  and Schwenk}}]{gezerlis2014pairing}
\bibinfo{author}{\bibfnamefont{A.}~\bibnamefont{Gezerlis}},
  \bibinfo{author}{\bibfnamefont{C.}~\bibnamefont{Pethick}}, \bibnamefont{and}
  \bibinfo{author}{\bibfnamefont{A.}~\bibnamefont{Schwenk}},
  \bibinfo{journal}{arXiv preprint arXiv:1406.6109}  (\bibinfo{year}{2014}).

\bibitem[{\citenamefont{Pastore et~al.}(2017)\citenamefont{Pastore, Shelley,
  Baroni, and Diget}}]{pastore2017new}
\bibinfo{author}{\bibfnamefont{A.}~\bibnamefont{Pastore}},
  \bibinfo{author}{\bibfnamefont{M.}~\bibnamefont{Shelley}},
  \bibinfo{author}{\bibfnamefont{S.}~\bibnamefont{Baroni}}, \bibnamefont{and}
  \bibinfo{author}{\bibfnamefont{C.}~\bibnamefont{Diget}},
  \bibinfo{journal}{Journal of Physics G: Nuclear and Particle Physics}
  \textbf{\bibinfo{volume}{44}}, \bibinfo{pages}{094003}
  (\bibinfo{year}{2017}).

\bibitem[{\citenamefont{Alford et~al.}(2007)\citenamefont{Alford, Blaschke,
  Drago, Kl{\"a}hn, Pagliara, and Schaffner-Bielich}}]{alford2007astrophysics}
\bibinfo{author}{\bibfnamefont{M.}~\bibnamefont{Alford}},
  \bibinfo{author}{\bibfnamefont{D.}~\bibnamefont{Blaschke}},
  \bibinfo{author}{\bibfnamefont{A.}~\bibnamefont{Drago}},
  \bibinfo{author}{\bibfnamefont{T.}~\bibnamefont{Kl{\"a}hn}},
  \bibinfo{author}{\bibfnamefont{G.}~\bibnamefont{Pagliara}}, \bibnamefont{and}
  \bibinfo{author}{\bibfnamefont{J.}~\bibnamefont{Schaffner-Bielich}},
  \bibinfo{journal}{Nature} \textbf{\bibinfo{volume}{445}}, \bibinfo{pages}{E7}
  (\bibinfo{year}{2007}).

\bibitem[{\citenamefont{Chamel and Haensel}(2008)}]{chamel2008physics}
\bibinfo{author}{\bibfnamefont{N.}~\bibnamefont{Chamel}} \bibnamefont{and}
  \bibinfo{author}{\bibfnamefont{P.}~\bibnamefont{Haensel}},
  \bibinfo{journal}{Living Reviews in relativity}
  \textbf{\bibinfo{volume}{11}}, \bibinfo{pages}{10} (\bibinfo{year}{2008}).

\bibitem[{\citenamefont{Sharma et~al.}(2015)\citenamefont{Sharma, Centelles,
  Vi{\~n}as, Baldo, and Burgio}}]{sharma2015unified}
\bibinfo{author}{\bibfnamefont{B.}~\bibnamefont{Sharma}},
  \bibinfo{author}{\bibfnamefont{M.}~\bibnamefont{Centelles}},
  \bibinfo{author}{\bibfnamefont{X.}~\bibnamefont{Vi{\~n}as}},
  \bibinfo{author}{\bibfnamefont{M.}~\bibnamefont{Baldo}}, \bibnamefont{and}
  \bibinfo{author}{\bibfnamefont{G.}~\bibnamefont{Burgio}},
  \bibinfo{journal}{Astronomy \& Astrophysics} \textbf{\bibinfo{volume}{584}},
  \bibinfo{pages}{A103} (\bibinfo{year}{2015}).

\bibitem[{\citenamefont{Chatterjee and Vidana}(2016)}]{chatterjee2016hyperons}
\bibinfo{author}{\bibfnamefont{D.}~\bibnamefont{Chatterjee}} \bibnamefont{and}
  \bibinfo{author}{\bibfnamefont{I.}~\bibnamefont{Vidana}},
  \bibinfo{journal}{The European Physical Journal A}
  \textbf{\bibinfo{volume}{52}}, \bibinfo{pages}{29} (\bibinfo{year}{2016}).

\bibitem[{\citenamefont{Vidana et~al.}(2018)\citenamefont{Vidana, Bashkanov,
  Watts, and Pastore}}]{vidana2018d}
\bibinfo{author}{\bibfnamefont{I.}~\bibnamefont{Vidana}},
  \bibinfo{author}{\bibfnamefont{M.}~\bibnamefont{Bashkanov}},
  \bibinfo{author}{\bibfnamefont{D.}~\bibnamefont{Watts}}, \bibnamefont{and}
  \bibinfo{author}{\bibfnamefont{A.}~\bibnamefont{Pastore}},
  \bibinfo{journal}{Physics Letters B} \textbf{\bibinfo{volume}{781}},
  \bibinfo{pages}{112} (\bibinfo{year}{2018}).

\bibitem[{\citenamefont{Baym et~al.}(1971)\citenamefont{Baym, Pethick, and
  Sutherland}}]{baym1971ground}
\bibinfo{author}{\bibfnamefont{G.}~\bibnamefont{Baym}},
  \bibinfo{author}{\bibfnamefont{C.}~\bibnamefont{Pethick}}, \bibnamefont{and}
  \bibinfo{author}{\bibfnamefont{P.}~\bibnamefont{Sutherland}},
  \bibinfo{journal}{The Astrophysical Journal} \textbf{\bibinfo{volume}{170}},
  \bibinfo{pages}{299} (\bibinfo{year}{1971}).

\bibitem[{\citenamefont{Dobaczewski et~al.}(2014)\citenamefont{Dobaczewski,
  Nazarewicz, and Reinhard}}]{dobaczewski2014error}
\bibinfo{author}{\bibfnamefont{J.}~\bibnamefont{Dobaczewski}},
  \bibinfo{author}{\bibfnamefont{W.}~\bibnamefont{Nazarewicz}},
  \bibnamefont{and} \bibinfo{author}{\bibfnamefont{P.}~\bibnamefont{Reinhard}},
  \bibinfo{journal}{Journal of Physics G: Nuclear and Particle Physics}
  \textbf{\bibinfo{volume}{41}}, \bibinfo{pages}{074001}
  (\bibinfo{year}{2014}).

\bibitem[{\citenamefont{Schunck et~al.}(2015)\citenamefont{Schunck, McDonnell,
  Sarich, Wild, and Higdon}}]{schunck2015error}
\bibinfo{author}{\bibfnamefont{N.}~\bibnamefont{Schunck}},
  \bibinfo{author}{\bibfnamefont{J.~D.} \bibnamefont{McDonnell}},
  \bibinfo{author}{\bibfnamefont{J.}~\bibnamefont{Sarich}},
  \bibinfo{author}{\bibfnamefont{S.~M.} \bibnamefont{Wild}}, \bibnamefont{and}
  \bibinfo{author}{\bibfnamefont{D.}~\bibnamefont{Higdon}},
  \bibinfo{journal}{Journal of Physics G: Nuclear and Particle Physics}
  \textbf{\bibinfo{volume}{42}}, \bibinfo{pages}{034024}
  (\bibinfo{year}{2015}).

\bibitem[{\citenamefont{McDonnell et~al.}(2015)\citenamefont{McDonnell,
  Schunck, Higdon, Sarich, Wild, and Nazarewicz}}]{mcdonnell2015uncertainty}
\bibinfo{author}{\bibfnamefont{J.}~\bibnamefont{McDonnell}},
  \bibinfo{author}{\bibfnamefont{N.}~\bibnamefont{Schunck}},
  \bibinfo{author}{\bibfnamefont{D.}~\bibnamefont{Higdon}},
  \bibinfo{author}{\bibfnamefont{J.}~\bibnamefont{Sarich}},
  \bibinfo{author}{\bibfnamefont{S.}~\bibnamefont{Wild}}, \bibnamefont{and}
  \bibinfo{author}{\bibfnamefont{W.}~\bibnamefont{Nazarewicz}},
  \bibinfo{journal}{Physical review letters} \textbf{\bibinfo{volume}{114}},
  \bibinfo{pages}{122501} (\bibinfo{year}{2015}).

\bibitem[{\citenamefont{Neufcourt et~al.}(2018)\citenamefont{Neufcourt, Cao,
  Nazarewicz, Viens et~al.}}]{neufcourt2018bayesian}
\bibinfo{author}{\bibfnamefont{L.}~\bibnamefont{Neufcourt}},
  \bibinfo{author}{\bibfnamefont{Y.}~\bibnamefont{Cao}},
  \bibinfo{author}{\bibfnamefont{W.}~\bibnamefont{Nazarewicz}},
  \bibinfo{author}{\bibfnamefont{F.}~\bibnamefont{Viens}},
  \bibnamefont{et~al.}, \bibinfo{journal}{Physical Review C}
  \textbf{\bibinfo{volume}{98}}, \bibinfo{pages}{034318}
  (\bibinfo{year}{2018}).

\bibitem[{\citenamefont{Utama and
  Piekarewicz}(2017{\natexlab{a}})}]{utama2017refining}
\bibinfo{author}{\bibfnamefont{R.}~\bibnamefont{Utama}} \bibnamefont{and}
  \bibinfo{author}{\bibfnamefont{J.}~\bibnamefont{Piekarewicz}},
  \bibinfo{journal}{Physical Review C} \textbf{\bibinfo{volume}{96}},
  \bibinfo{pages}{044308} (\bibinfo{year}{2017}{\natexlab{a}}).

\bibitem[{\citenamefont{Utama and Piekarewicz}(2018)}]{utama2018validating}
\bibinfo{author}{\bibfnamefont{R.}~\bibnamefont{Utama}} \bibnamefont{and}
  \bibinfo{author}{\bibfnamefont{J.}~\bibnamefont{Piekarewicz}},
  \bibinfo{journal}{Physical Review C} \textbf{\bibinfo{volume}{97}},
  \bibinfo{pages}{014306} (\bibinfo{year}{2018}).

\bibitem[{\citenamefont{Sprouse et~al.}(2019)\citenamefont{Sprouse, Perez,
  Surman, Mumpower, McLaughlin, and Schunck}}]{sprouse2019propagation}
\bibinfo{author}{\bibfnamefont{T.}~\bibnamefont{Sprouse}},
  \bibinfo{author}{\bibfnamefont{R.~N.} \bibnamefont{Perez}},
  \bibinfo{author}{\bibfnamefont{R.}~\bibnamefont{Surman}},
  \bibinfo{author}{\bibfnamefont{M.}~\bibnamefont{Mumpower}},
  \bibinfo{author}{\bibfnamefont{G.}~\bibnamefont{McLaughlin}},
  \bibnamefont{and} \bibinfo{author}{\bibfnamefont{N.}~\bibnamefont{Schunck}},
  \bibinfo{journal}{arXiv preprint arXiv:1901.10337}  (\bibinfo{year}{2019}).

\bibitem[{\citenamefont{Kortelainen}(2015)}]{kortelainen2015propagation}
\bibinfo{author}{\bibfnamefont{M.}~\bibnamefont{Kortelainen}},
  \bibinfo{journal}{Journal of Physics G: Nuclear and Particle Physics}
  \textbf{\bibinfo{volume}{42}}, \bibinfo{pages}{034021}
  (\bibinfo{year}{2015}).

\bibitem[{\citenamefont{Roca-Maza et~al.}(2015)\citenamefont{Roca-Maza, Paar,
  and Colo}}]{roca2015covariance}
\bibinfo{author}{\bibfnamefont{X.}~\bibnamefont{Roca-Maza}},
  \bibinfo{author}{\bibfnamefont{N.}~\bibnamefont{Paar}}, \bibnamefont{and}
  \bibinfo{author}{\bibfnamefont{G.}~\bibnamefont{Colo}},
  \bibinfo{journal}{Journal of Physics G: Nuclear and Particle Physics}
  \textbf{\bibinfo{volume}{42}}, \bibinfo{pages}{034033}
  (\bibinfo{year}{2015}).

\bibitem[{\citenamefont{Becker et~al.}(2018)\citenamefont{Becker, Pastore,
  Davesne, and Navarro}}]{becker2018error}
\bibinfo{author}{\bibfnamefont{P.}~\bibnamefont{Becker}},
  \bibinfo{author}{\bibfnamefont{A.}~\bibnamefont{Pastore}},
  \bibinfo{author}{\bibfnamefont{D.}~\bibnamefont{Davesne}}, \bibnamefont{and}
  \bibinfo{author}{\bibfnamefont{J.}~\bibnamefont{Navarro}},
  \bibinfo{journal}{arXiv preprint arXiv:1811.07866}  (\bibinfo{year}{2018}).

\bibitem[{\citenamefont{Duflo and Zuker}(1995)}]{duflo1995microscopic}
\bibinfo{author}{\bibfnamefont{J.}~\bibnamefont{Duflo}} \bibnamefont{and}
  \bibinfo{author}{\bibfnamefont{A.}~\bibnamefont{Zuker}},
  \bibinfo{journal}{Physical Review C} \textbf{\bibinfo{volume}{52}},
  \bibinfo{pages}{R23} (\bibinfo{year}{1995}).

\bibitem[{\citenamefont{Mendoza-Temis et~al.}(2010)\citenamefont{Mendoza-Temis,
  Hirsch, and Zuker}}]{mendoza2010anatomy}
\bibinfo{author}{\bibfnamefont{J.}~\bibnamefont{Mendoza-Temis}},
  \bibinfo{author}{\bibfnamefont{J.~G.} \bibnamefont{Hirsch}},
  \bibnamefont{and} \bibinfo{author}{\bibfnamefont{A.~P.} \bibnamefont{Zuker}},
  \bibinfo{journal}{Nuclear Physics A} \textbf{\bibinfo{volume}{843}},
  \bibinfo{pages}{14} (\bibinfo{year}{2010}).

\bibitem[{\citenamefont{Zuker}(2011)}]{zuker2011anatomy}
\bibinfo{author}{\bibfnamefont{A.}~\bibnamefont{Zuker}}, in
  \emph{\bibinfo{booktitle}{11th Symposium on Nuclei in the Cosmos}}
  (\bibinfo{organization}{SISSA Medialab}, \bibinfo{year}{2011}), vol.
  \bibinfo{volume}{100}, p. \bibinfo{pages}{083}.

\bibitem[{\citenamefont{Qi}(2015)}]{qi2015theoretical}
\bibinfo{author}{\bibfnamefont{C.}~\bibnamefont{Qi}}, \bibinfo{journal}{Journal
  of Physics G: Nuclear and Particle Physics} \textbf{\bibinfo{volume}{42}},
  \bibinfo{pages}{045104} (\bibinfo{year}{2015}).

\bibitem[{\citenamefont{Pastore}(2019)}]{pastore2019introduction}
\bibinfo{author}{\bibfnamefont{A.}~\bibnamefont{Pastore}},
  \bibinfo{journal}{Journal of Physics G: Nuclear and Particle Physics}
  \textbf{\bibinfo{volume}{46}}, \bibinfo{pages}{052001}
  (\bibinfo{year}{2019}).

\bibitem[{\citenamefont{Wang et~al.}(2017)\citenamefont{Wang, Audi, Kondev,
  Huang, Naimi, and Xu}}]{wang2017ame2016}
\bibinfo{author}{\bibfnamefont{M.}~\bibnamefont{Wang}},
  \bibinfo{author}{\bibfnamefont{G.}~\bibnamefont{Audi}},
  \bibinfo{author}{\bibfnamefont{F.}~\bibnamefont{Kondev}},
  \bibinfo{author}{\bibfnamefont{W.}~\bibnamefont{Huang}},
  \bibinfo{author}{\bibfnamefont{S.}~\bibnamefont{Naimi}}, \bibnamefont{and}
  \bibinfo{author}{\bibfnamefont{X.}~\bibnamefont{Xu}},
  \bibinfo{journal}{Chinese Physics C} \textbf{\bibinfo{volume}{41}},
  \bibinfo{pages}{030003} (\bibinfo{year}{2017}).

\bibitem[{\citenamefont{Vetterling et~al.}(1994)\citenamefont{Vetterling,
  Teukolsky, Press, and Flannery}}]{vetterling1994numerical}
\bibinfo{author}{\bibfnamefont{W.~T.} \bibnamefont{Vetterling}},
  \bibinfo{author}{\bibfnamefont{S.~A.} \bibnamefont{Teukolsky}},
  \bibinfo{author}{\bibfnamefont{W.~H.} \bibnamefont{Press}}, \bibnamefont{and}
  \bibinfo{author}{\bibfnamefont{B.~P.} \bibnamefont{Flannery}},
  \emph{\bibinfo{title}{Numerical recipes}}, vol.~\bibinfo{volume}{3}
  (\bibinfo{publisher}{Cambridge University Press Cambridge:},
  \bibinfo{year}{1994}).

\bibitem[{\citenamefont{Audi and Wapstra}(1995)}]{audi19951995}
\bibinfo{author}{\bibfnamefont{G.}~\bibnamefont{Audi}} \bibnamefont{and}
  \bibinfo{author}{\bibfnamefont{A.}~\bibnamefont{Wapstra}},
  \bibinfo{journal}{Nuclear Physics A} \textbf{\bibinfo{volume}{595}},
  \bibinfo{pages}{409} (\bibinfo{year}{1995}).

\bibitem[{\citenamefont{Utama and Piekarewicz}(2017{\natexlab{b}})}]{uta17}
\bibinfo{author}{\bibfnamefont{R.}~\bibnamefont{Utama}} \bibnamefont{and}
  \bibinfo{author}{\bibfnamefont{J.}~\bibnamefont{Piekarewicz}},
  \bibinfo{journal}{Physical Review C} \textbf{\bibinfo{volume}{96}},
  \bibinfo{pages}{044308} (\bibinfo{year}{2017}{\natexlab{b}}).

\bibitem[{\citenamefont{Wang et~al.}(2012)\citenamefont{Wang, Audi, Wapstra,
  Kondev, MacCormick, Xu, and Pfeiffer}}]{wang2012ame2012}
\bibinfo{author}{\bibfnamefont{M.}~\bibnamefont{Wang}},
  \bibinfo{author}{\bibfnamefont{G.}~\bibnamefont{Audi}},
  \bibinfo{author}{\bibfnamefont{A.}~\bibnamefont{Wapstra}},
  \bibinfo{author}{\bibfnamefont{F.}~\bibnamefont{Kondev}},
  \bibinfo{author}{\bibfnamefont{M.}~\bibnamefont{MacCormick}},
  \bibinfo{author}{\bibfnamefont{X.}~\bibnamefont{Xu}}, \bibnamefont{and}
  \bibinfo{author}{\bibfnamefont{B.}~\bibnamefont{Pfeiffer}},
  \bibinfo{journal}{Chinese Physics C} \textbf{\bibinfo{volume}{36}},
  \bibinfo{pages}{1603} (\bibinfo{year}{2012}).

\bibitem[{\citenamefont{R.J.Barlow}(1989)}]{bar89}
\bibinfo{author}{\bibnamefont{R.J.Barlow}}, \emph{\bibinfo{title}{A Guide to
  the Use of Statistical Methods in the Physical Sciences}}
  (\bibinfo{publisher}{John Wiley}, \bibinfo{year}{1989}).

\bibitem[{\citenamefont{Bertsch and Bingham}(2017)}]{bertsch2017estimating}
\bibinfo{author}{\bibfnamefont{G.}~\bibnamefont{Bertsch}} \bibnamefont{and}
  \bibinfo{author}{\bibfnamefont{D.}~\bibnamefont{Bingham}},
  \bibinfo{journal}{Physical review letters} \textbf{\bibinfo{volume}{119}},
  \bibinfo{pages}{252501} (\bibinfo{year}{2017}).

\bibitem[{\citenamefont{Brockwell and Davis}(2013)}]{brockwell2013time}
\bibinfo{author}{\bibfnamefont{P.~J.} \bibnamefont{Brockwell}}
  \bibnamefont{and} \bibinfo{author}{\bibfnamefont{R.~A.} \bibnamefont{Davis}},
  \emph{\bibinfo{title}{Time series: theory and methods}}
  (\bibinfo{publisher}{Springer Science \& Business Media},
  \bibinfo{year}{2013}).

\bibitem[{\citenamefont{Efron and Tibshirani}(1994)}]{efron1994introduction}
\bibinfo{author}{\bibfnamefont{B.}~\bibnamefont{Efron}} \bibnamefont{and}
  \bibinfo{author}{\bibfnamefont{R.~J.} \bibnamefont{Tibshirani}},
  \emph{\bibinfo{title}{An introduction to the bootstrap}}
  (\bibinfo{publisher}{CRC press}, \bibinfo{year}{1994}).

\bibitem[{\citenamefont{Kreiss and Paparoditis}(2011)}]{kreiss2011bootstrap}
\bibinfo{author}{\bibfnamefont{J.-P.} \bibnamefont{Kreiss}} \bibnamefont{and}
  \bibinfo{author}{\bibfnamefont{E.}~\bibnamefont{Paparoditis}},
  \bibinfo{journal}{Journal of the Korean Statistical Society}
  \textbf{\bibinfo{volume}{40}}, \bibinfo{pages}{357} (\bibinfo{year}{2011}).

\bibitem[{\citenamefont{Chernick et~al.}(2011)\citenamefont{Chernick,
  Gonz{\'a}lez-Manteiga, Crujeiras, and Barrios}}]{chernick2011bootstrap}
\bibinfo{author}{\bibfnamefont{M.~R.} \bibnamefont{Chernick}},
  \bibinfo{author}{\bibfnamefont{W.}~\bibnamefont{Gonz{\'a}lez-Manteiga}},
  \bibinfo{author}{\bibfnamefont{R.~M.} \bibnamefont{Crujeiras}},
  \bibnamefont{and} \bibinfo{author}{\bibfnamefont{E.~B.}
  \bibnamefont{Barrios}}, in \emph{\bibinfo{booktitle}{International
  encyclopedia of statistical science}} (\bibinfo{publisher}{Springer},
  \bibinfo{year}{2011}), pp. \bibinfo{pages}{169--174}.

\bibitem[{\citenamefont{Muir et~al.}(2018)\citenamefont{Muir, Pastore,
  Dobaczewski, and Barton}}]{muir2018bootstrap}
\bibinfo{author}{\bibfnamefont{D.}~\bibnamefont{Muir}},
  \bibinfo{author}{\bibfnamefont{A.}~\bibnamefont{Pastore}},
  \bibinfo{author}{\bibfnamefont{J.}~\bibnamefont{Dobaczewski}},
  \bibnamefont{and} \bibinfo{author}{\bibfnamefont{C.}~\bibnamefont{Barton}},
  \bibinfo{journal}{Acta Physica Polonica B} \textbf{\bibinfo{volume}{49}}
  (\bibinfo{year}{2018}).

\bibitem[{\citenamefont{Efron}(1979)}]{efr79}
\bibinfo{author}{\bibfnamefont{B.}~\bibnamefont{Efron}},
  \bibinfo{journal}{Annals of Statistics} \textbf{\bibinfo{volume}{7}},
  \bibinfo{pages}{1} (\bibinfo{year}{1979}).

\bibitem[{\citenamefont{Rasmussen}(1987)}]{ras87}
\bibinfo{author}{\bibfnamefont{J.~L.} \bibnamefont{Rasmussen}},
  \bibinfo{journal}{Psychological Bulletin} \textbf{\bibinfo{volume}{101}},
  \bibinfo{pages}{136} (\bibinfo{year}{1987}).

\bibitem[{\citenamefont{Fisher and Hall}(1990)}]{fisher1990new}
\bibinfo{author}{\bibfnamefont{N.~I.} \bibnamefont{Fisher}} \bibnamefont{and}
  \bibinfo{author}{\bibfnamefont{P.}~\bibnamefont{Hall}},
  \bibinfo{journal}{Geophysical Journal International}
  \textbf{\bibinfo{volume}{101}}, \bibinfo{pages}{305} (\bibinfo{year}{1990}).

\bibitem[{\citenamefont{Sauerbrei and
  Schumacher}(1992)}]{sauerbrei1992bootstrap}
\bibinfo{author}{\bibfnamefont{W.}~\bibnamefont{Sauerbrei}} \bibnamefont{and}
  \bibinfo{author}{\bibfnamefont{M.}~\bibnamefont{Schumacher}},
  \bibinfo{journal}{Statistics in medicine} \textbf{\bibinfo{volume}{11}},
  \bibinfo{pages}{2093} (\bibinfo{year}{1992}).

\bibitem[{\citenamefont{Manly}(2006)}]{manly2006randomization}
\bibinfo{author}{\bibfnamefont{B.~F.} \bibnamefont{Manly}},
  vol.~\bibinfo{volume}{70} (\bibinfo{publisher}{CRC press},
  \bibinfo{year}{2006}).

\bibitem[{\citenamefont{P{\'e}rez et~al.}(2014)\citenamefont{P{\'e}rez, Amaro,
  and Arriola}}]{per14}
\bibinfo{author}{\bibfnamefont{R.~N.} \bibnamefont{P{\'e}rez}},
  \bibinfo{author}{\bibfnamefont{J.}~\bibnamefont{Amaro}}, \bibnamefont{and}
  \bibinfo{author}{\bibfnamefont{E.~R.} \bibnamefont{Arriola}},
  \bibinfo{journal}{Physics Letters B} \textbf{\bibinfo{volume}{738}},
  \bibinfo{pages}{155} (\bibinfo{year}{2014}).

\bibitem[{\citenamefont{Pasquini et~al.}(2018)\citenamefont{Pasquini, Pedroni,
  and Sconfietti}}]{pasq18}
\bibinfo{author}{\bibfnamefont{B.}~\bibnamefont{Pasquini}},
  \bibinfo{author}{\bibfnamefont{P.}~\bibnamefont{Pedroni}}, \bibnamefont{and}
  \bibinfo{author}{\bibfnamefont{S.}~\bibnamefont{Sconfietti}},
  \bibinfo{journal}{Physical Review C} \textbf{\bibinfo{volume}{98}},
  \bibinfo{pages}{015204} (\bibinfo{year}{2018}).

\bibitem[{\citenamefont{Kreiss and Lahiri}(2012)}]{kreiss2012bootstrap}
\bibinfo{author}{\bibfnamefont{J.-P.} \bibnamefont{Kreiss}} \bibnamefont{and}
  \bibinfo{author}{\bibfnamefont{S.~N.} \bibnamefont{Lahiri}}, in
  \emph{\bibinfo{booktitle}{Handbook of statistics}}
  (\bibinfo{publisher}{Elsevier}, \bibinfo{year}{2012}),
  vol.~\bibinfo{volume}{30}, pp. \bibinfo{pages}{3--26}.

\bibitem[{\citenamefont{Nik{\v{s}}i{\'c} and
  Vretenar}(2016)}]{nikvsic2016sloppy}
\bibinfo{author}{\bibfnamefont{T.}~\bibnamefont{Nik{\v{s}}i{\'c}}}
  \bibnamefont{and} \bibinfo{author}{\bibfnamefont{D.}~\bibnamefont{Vretenar}},
  \bibinfo{journal}{Physical Review C} \textbf{\bibinfo{volume}{94}},
  \bibinfo{pages}{024333} (\bibinfo{year}{2016}).

\bibitem[{\citenamefont{Gurney}(2014)}]{gurney2014introduction}
\bibinfo{author}{\bibfnamefont{K.}~\bibnamefont{Gurney}},
  \emph{\bibinfo{title}{An introduction to neural networks}}
  (\bibinfo{publisher}{CRC press}, \bibinfo{year}{2014}).

\bibitem[{\citenamefont{Svozil et~al.}(1997)\citenamefont{Svozil, Kvasnicka,
  and Pospichal}}]{svozil1997introduction}
\bibinfo{author}{\bibfnamefont{D.}~\bibnamefont{Svozil}},
  \bibinfo{author}{\bibfnamefont{V.}~\bibnamefont{Kvasnicka}},
  \bibnamefont{and}
  \bibinfo{author}{\bibfnamefont{J.}~\bibnamefont{Pospichal}},
  \bibinfo{journal}{Chemometrics and intelligent laboratory systems}
  \textbf{\bibinfo{volume}{39}}, \bibinfo{pages}{43} (\bibinfo{year}{1997}).

\bibitem[{\citenamefont{Murata et~al.}(1994)\citenamefont{Murata, Yoshizawa,
  and Amari}}]{murata1994network}
\bibinfo{author}{\bibfnamefont{N.}~\bibnamefont{Murata}},
  \bibinfo{author}{\bibfnamefont{S.}~\bibnamefont{Yoshizawa}},
  \bibnamefont{and} \bibinfo{author}{\bibfnamefont{S.-i.} \bibnamefont{Amari}},
  \bibinfo{journal}{IEEE transactions on neural networks}
  \textbf{\bibinfo{volume}{5}}, \bibinfo{pages}{865} (\bibinfo{year}{1994}).

\bibitem[{\citenamefont{Niu and Liang}(2018)}]{niu2018nuclear}
\bibinfo{author}{\bibfnamefont{Z.}~\bibnamefont{Niu}} \bibnamefont{and}
  \bibinfo{author}{\bibfnamefont{H.}~\bibnamefont{Liang}},
  \bibinfo{journal}{Physics Letters B} \textbf{\bibinfo{volume}{778}},
  \bibinfo{pages}{48} (\bibinfo{year}{2018}).

\bibitem[{\citenamefont{Welker et~al.}(2017)\citenamefont{Welker, Althubiti,
  Atanasov, Blaum, Cocolios, Herfurth, Kreim, Lunney, Manea, Mougeot
  et~al.}}]{welker2017binding}
\bibinfo{author}{\bibfnamefont{A.}~\bibnamefont{Welker}},
  \bibinfo{author}{\bibfnamefont{N.}~\bibnamefont{Althubiti}},
  \bibinfo{author}{\bibfnamefont{D.}~\bibnamefont{Atanasov}},
  \bibinfo{author}{\bibfnamefont{K.}~\bibnamefont{Blaum}},
  \bibinfo{author}{\bibfnamefont{T.~E.} \bibnamefont{Cocolios}},
  \bibinfo{author}{\bibfnamefont{F.}~\bibnamefont{Herfurth}},
  \bibinfo{author}{\bibfnamefont{S.}~\bibnamefont{Kreim}},
  \bibinfo{author}{\bibfnamefont{D.}~\bibnamefont{Lunney}},
  \bibinfo{author}{\bibfnamefont{V.}~\bibnamefont{Manea}},
  \bibinfo{author}{\bibfnamefont{M.}~\bibnamefont{Mougeot}},
  \bibnamefont{et~al.}, \bibinfo{journal}{Physical review letters}
  \textbf{\bibinfo{volume}{119}}, \bibinfo{pages}{192502}
  (\bibinfo{year}{2017}).

\bibitem[{\citenamefont{Idini}(2019)}]{idini2019pac}
\bibinfo{author}{\bibfnamefont{A.}~\bibnamefont{Idini}},
  \bibinfo{journal}{arXiv preprint arXiv:1904.00057}  (\bibinfo{year}{2019}).

\bibitem[{\citenamefont{Barea et~al.}(2005)\citenamefont{Barea, Frank, Hirsch,
  and Van~Isacker}}]{barea2005nuclear}
\bibinfo{author}{\bibfnamefont{J.}~\bibnamefont{Barea}},
  \bibinfo{author}{\bibfnamefont{A.}~\bibnamefont{Frank}},
  \bibinfo{author}{\bibfnamefont{J.~G.} \bibnamefont{Hirsch}},
  \bibnamefont{and}
  \bibinfo{author}{\bibfnamefont{P.}~\bibnamefont{Van~Isacker}},
  \bibinfo{journal}{Physical review letters} \textbf{\bibinfo{volume}{94}},
  \bibinfo{pages}{102501} (\bibinfo{year}{2005}).

\bibitem[{\citenamefont{Roca-Maza and Piekarewicz}(2008)}]{roca2008impact}
\bibinfo{author}{\bibfnamefont{X.}~\bibnamefont{Roca-Maza}} \bibnamefont{and}
  \bibinfo{author}{\bibfnamefont{J.}~\bibnamefont{Piekarewicz}},
  \bibinfo{journal}{Physical Review C} \textbf{\bibinfo{volume}{78}},
  \bibinfo{pages}{025807} (\bibinfo{year}{2008}).

\bibitem[{\citenamefont{Coldwell-Horsfall and
  Maradudin}(1960)}]{coldwell1960zero}
\bibinfo{author}{\bibfnamefont{R.~A.} \bibnamefont{Coldwell-Horsfall}}
  \bibnamefont{and} \bibinfo{author}{\bibfnamefont{A.~A.}
  \bibnamefont{Maradudin}}, \bibinfo{journal}{Journal of Mathematical Physics}
  \textbf{\bibinfo{volume}{1}}, \bibinfo{pages}{395} (\bibinfo{year}{1960}).

\bibitem[{\citenamefont{R{\"u}ster et~al.}(2006)\citenamefont{R{\"u}ster,
  Hempel, and Schaffner-Bielich}}]{ruster2006outer}
\bibinfo{author}{\bibfnamefont{S.~B.} \bibnamefont{R{\"u}ster}},
  \bibinfo{author}{\bibfnamefont{M.}~\bibnamefont{Hempel}}, \bibnamefont{and}
  \bibinfo{author}{\bibfnamefont{J.}~\bibnamefont{Schaffner-Bielich}},
  \bibinfo{journal}{Physical Review C} \textbf{\bibinfo{volume}{73}},
  \bibinfo{pages}{035804} (\bibinfo{year}{2006}).

\bibitem[{\citenamefont{Pearson et~al.}(2011)\citenamefont{Pearson, Goriely,
  and Chamel}}]{pearson2011properties}
\bibinfo{author}{\bibfnamefont{J.}~\bibnamefont{Pearson}},
  \bibinfo{author}{\bibfnamefont{S.}~\bibnamefont{Goriely}}, \bibnamefont{and}
  \bibinfo{author}{\bibfnamefont{N.}~\bibnamefont{Chamel}},
  \bibinfo{journal}{Physical Review C} \textbf{\bibinfo{volume}{83}},
  \bibinfo{pages}{065810} (\bibinfo{year}{2011}).

\bibitem[{\citenamefont{Sobiczewski and
  Litvinov}(2014)}]{sobiczewski2014predictive}
\bibinfo{author}{\bibfnamefont{A.}~\bibnamefont{Sobiczewski}} \bibnamefont{and}
  \bibinfo{author}{\bibfnamefont{Y.~A.} \bibnamefont{Litvinov}},
  \bibinfo{journal}{Physical Review C} \textbf{\bibinfo{volume}{90}},
  \bibinfo{pages}{017302} (\bibinfo{year}{2014}).

\bibitem[{\citenamefont{Neill et~al.}(2019)\citenamefont{Neill, Medler,
  Pastore, and Barton}}]{neill2019impact}
\bibinfo{author}{\bibfnamefont{D.}~\bibnamefont{Neill}},
  \bibinfo{author}{\bibfnamefont{K.}~\bibnamefont{Medler}},
  \bibinfo{author}{\bibfnamefont{A.}~\bibnamefont{Pastore}}, \bibnamefont{and}
  \bibinfo{author}{\bibfnamefont{C.}~\bibnamefont{Barton}},
  \bibinfo{journal}{arXiv preprint arXiv:1909.11496}  (\bibinfo{year}{2019}).

\bibitem[{\citenamefont{Kirson}(2012)}]{kirson2012empirical}
\bibinfo{author}{\bibfnamefont{M.~W.} \bibnamefont{Kirson}},
  \bibinfo{journal}{Nuclear Physics A} \textbf{\bibinfo{volume}{893}},
  \bibinfo{pages}{27} (\bibinfo{year}{2012}).

\bibitem[{\citenamefont{Fortin et~al.}(2016)\citenamefont{Fortin,
  Provid{\^e}ncia, Raduta, Gulminelli, Zdunik, Haensel, and
  Bejger}}]{fortin2016neutron}
\bibinfo{author}{\bibfnamefont{M.}~\bibnamefont{Fortin}},
  \bibinfo{author}{\bibfnamefont{C.}~\bibnamefont{Provid{\^e}ncia}},
  \bibinfo{author}{\bibfnamefont{A.~R.} \bibnamefont{Raduta}},
  \bibinfo{author}{\bibfnamefont{F.}~\bibnamefont{Gulminelli}},
  \bibinfo{author}{\bibfnamefont{J.}~\bibnamefont{Zdunik}},
  \bibinfo{author}{\bibfnamefont{P.}~\bibnamefont{Haensel}}, \bibnamefont{and}
  \bibinfo{author}{\bibfnamefont{M.}~\bibnamefont{Bejger}},
  \bibinfo{journal}{Physical Review C} \textbf{\bibinfo{volume}{94}},
  \bibinfo{pages}{035804} (\bibinfo{year}{2016}).

\bibitem[{\citenamefont{Liu et~al.}(2011)\citenamefont{Liu, Wang, Deng, and
  Wu}}]{min11}
\bibinfo{author}{\bibfnamefont{M.}~\bibnamefont{Liu}},
  \bibinfo{author}{\bibfnamefont{N.}~\bibnamefont{Wang}},
  \bibinfo{author}{\bibfnamefont{Y.}~\bibnamefont{Deng}}, \bibnamefont{and}
  \bibinfo{author}{\bibfnamefont{X.}~\bibnamefont{Wu}}, \bibinfo{journal}{Phys.
  Rev. C} \textbf{\bibinfo{volume}{84}}, \bibinfo{pages}{014333}
  (\bibinfo{year}{2011}),
  \urlprefix\url{https://link.aps.org/doi/10.1103/PhysRevC.84.014333}.

\end{thebibliography}

\end{document}